# Scaling approaches to steady wall-induced turbulence

Paul C. Fife

August 12, 2006

**Abstract**

The problem of discerning key features of steady turbulent flow adjacent to a wall has drawn the attention of some of the most noted fluid dynamicists of all time. Standard examples of such features are found in the mean velocity profiles of turbulent flow in channels, pipes or boundary layers. The aim of this review article is to expound the essence of some elementary theoretical efforts in this regard. Possibly the best known of them, and certainly the simplest, is the argument (obtained independently) by Izakson (1937) and Millikan (1939). They showed that if an inner scaling and an outer scaling for the profile are valid near the wall and near the center of the flow (or the edge of the boundary layer), respectively, and if there is an overlap region where both scalings are valid, then the profile must be logarithmic in that common region. That theoretical justification has been used and expanded upon by innumerable authors for over 60 years, and at the present time is still rightly enjoying popularity.

Although background discussions of several related topics are included in the present article, for example the classical ideas of Prandtl and von Karman, the main foci will be on (*i*) a careful examination of the Izakson-Millikan argument, together with a presentation of a better mathematical justification for its conclusion; and (*ii*) a detailed clarification of a newer approach to gaining theoretical understanding of the mean velocity and Reynolds stress profiles based on the "search for scaling patches".

The two approaches share common goals, they are both heavily involved with scaling concepts, and many results are similar, but the logical trains of thought are entirely different. The first, as mentioned, dates back to the 30's and the second was introduced in a series of recent papers by Fife et al, Wei et al, and Klewicki et al.

Our emphasis will be on the question of how and how well these arguments supply insight into the structure of the mean flow profiles. Although empirical results may initiate the search for explanations, they will be viewed simply as means to that end.

# Contents

# 1 Introduction

Every theoretical investigation of highly turbulent fluid dynamics is necessarily incomplete, because accepted accurate models such as the Navier-Stokes equations lie beyond the scope of full solution by existing methods, whether those methods be numerical or analytical.

Faced with this failing, researchers have often turned to seeking theoretical information through partial analyses, incomplete models, or reasoning which is not fully based on rigorous deduction. The search for such approaches is bound to yield, and historically has yielded, some fruitful avenues leading to insightful, though perhaps tentative, conclusions.

The goals of this paper are to examine, and elaborate on, some of the major elementary attempts in this vein to gain some understanding of the mechanisms behind steady (in the mean) turbulent incompressible flow bounded by a wall, as well as to expound some recent developments. The prototypical examples will be rapid flow in a channel or pipe forced by an imposed pressure gradient or the differential motion of the walls, and turbulent boundary layers of various types. There has been an enormous amount of work along these lines, and a complete review of it will not be given. Rather we look only at some efforts to gain insight into these mechanisms as they affect simple mean flow quantities, through the use of scaling concepts and averaged Navier-Stokes equations.

These efforts have been diverse, and when compared one with another have sometimes relied on assumptions which appear to be incompatible. Nevertheless despite any seeming incongruities, the various approaches should not necessarily be thought of as competing among themselves. Rather by contributing their individual insights, they collectively add to our understanding. No one approach can ever rationally be viewed as the last word on the subject.

So to summarize, the goal in these notes is to understand what phenomena relating to average events occur in wall-bounded turbulent flow and, more importantly, *why* they occur. This means that our approach will focus on supplying theoretical explanations for observed features of the flow, as well as providing predictions of features which can then be correlated with empirical data. When possible, we will not be content with pointing to experimental evidence for properties of the flow, but will strive to develop reasons why those properties should hold. These reasons will not generally be rigorous, and at times, of necessity, they will be partly based on empirical findings. But to advance the goal of basic understanding, they will, as far as possible, be grounded in theory, i.e. in accepted mathematical models. Mostly, these models will be built from the averaged Navier-Stokes equations.

## 2 The Navier-Stokes equations and Reynolds averaging

Our mathematical models use the symbols $\boldsymbol{u} = (u_1, u_2, u_3)$, $p$, $t$ and $\boldsymbol{x} = (x_1, x_2, x_3)$ to denote the velocity, pressure, time and space variables, with $\mu$ and $\rho$ the material constants of viscosity and fluid density.

The standard incompressible Navier-Stokes equations when there is no body force can be written as follows:
$$\frac{\partial \boldsymbol{u}}{\partial t} + (\boldsymbol{u} \cdot \nabla)\boldsymbol{u} = \nu \nabla^2 \boldsymbol{u} - \frac{1}{\rho}\nabla p, \tag{1}$$
where $\nu = \mu/\rho$ is the *kinematic viscosity*; and
$$\operatorname{div} \boldsymbol{u} = 0. \tag{2}$$
The equations (1) (a vector equation which has three components) and (2) are four equations in all, for four unknown functions: the three components of $\boldsymbol{u}$ plus the pressure $p$.

These equations are almost universally recognized as an accurate representation, on the macroscopic level, of flows of many standard kinds of incompressible fluids. Generalizations exist in many forms. The second (nonlinear) term on the left of (1) is called the "inertia term", and the first term on the right is the "viscosity term". Broad features of the solutions are typically governed by the order of magnitude of the Reynolds number, a ratio of the effects of these terms.

Here we start with a very common framework for studying turbulent motions called Reynolds averaging [1]. It is a short cut, and as such does not supply all the desired information about any given flow scenario. A wealth of details about fluid motions is erased. The modeler is left with the task of partially filling the resulting deficiency with assumptions about features of the fluid motion in its disorganized state. These attempts are bound to be hit or miss to some extent, but as we will see, there may be tests which one can apply to gauge the validity of such assumptions. For example, two or more avenues to obtaining models may yield similar results, which would tend to corroborate both. A very common remedy for the underdetermined nature of the Reynolds averaged equations is to posit closure relations. That approach is especially useful in numerical simulations. But the magnitude of these endeavors precludes their inclusion here.

The goal, then, is to use the Navier-Stokes equations in their averaged form, together with scaling considerations, to gain insight into some basic properties of wall-induced turbulence. Specifically, we focus on features of the mean velocity and Reynolds stress profiles, as functions of distance from the wall. The Reynolds stress is one of the measures of turbulence activity. There have in fact been several distinct approaches to the task of using the averaged equations to provide insight into wall-induced turbulence. That fact, together with the apparent simplicity of the formulation within that framework, makes the latter an ideal arena in which to explore the interaction of mathematical and intuitive tools to tackle a very complex problem.

We now review the well-known ideas behind the process of Reynolds averaging [1]. One supposes that at each spatial location $\boldsymbol{x}$ and time $t$, the velocity and the other flow quantities have well defined average values. This could be an imagined "ensemble" average over many similar observations, or over some small (but not too small) region in space-time containing the point in question. In the case of "steady" fully developed turbulence, it could be the time average. The mean (average) values may then depend on space and time, but on scales perhaps larger than those over which the average is taken.

Thus, we can write the velocity vector as its mean plus its fluctuation about the mean:

$$\boldsymbol{u}(\boldsymbol{x},t) = \boldsymbol{U}(\boldsymbol{x},t) + \boldsymbol{u}'(\boldsymbol{x},t), \tag{3}$$

with a similar decomposition for vorticity, pressure, etc. Averages will be taken of not only primary flow variables, but also of products of them. The averaging operation is denoted by angle brackets—for example $\langle \boldsymbol{u} \rangle = \boldsymbol{U}$; $\langle \partial_i \boldsymbol{u} \rangle = \partial_i \boldsymbol{U}$ for any derivative $\partial_i = \frac{\partial}{\partial x_i}$ or $\frac{\partial}{\partial t}$; $\langle \boldsymbol{u}' \rangle = 0$; $\langle U_i u'_j \rangle = 0$.

We substitute (3) into the Navier-Stokes equations (1), (2) and then take the average of the resulting equations to obtain

$$\partial_t \boldsymbol{U} - \nu \nabla^2 \boldsymbol{U} + (\boldsymbol{U} \cdot \nabla)\boldsymbol{U} + \langle \boldsymbol{u}' \cdot \nabla \boldsymbol{u}' \rangle + \frac{1}{\rho} \nabla P = 0, \tag{4}$$

$$\nabla \cdot \boldsymbol{U} = 0. \tag{5}$$

Note also that

$$\nabla \cdot \boldsymbol{u}' = 0. \tag{6}$$

The $i$-th component of the 4th term in (4) is

$$\langle u'_j \partial_j u'_i \rangle = \partial_j \langle u'_i u'_j \rangle$$

(summing over the repeated index), the equality by virtue of (6). This shows that the 4th term, being a divergence, acts as a pseudo-stress gradient. Accordingly, we define the "Reynolds stress tensor" $\boldsymbol{\tau}$ by

$$\tau_{ij} = -\langle u'_i u'_j \rangle, \tag{7}$$

which is symmetric, as any stress tensor should be. Therefore (4) becomes

$$\partial_t \boldsymbol{U} - \nu \nabla^2 \boldsymbol{U} + (\boldsymbol{U} \cdot \nabla)\boldsymbol{U} + \nabla \left( \frac{1}{\rho} P - \boldsymbol{\tau} \right) = 0, \tag{8}$$

where the last term can also be written $\nabla \cdot \boldsymbol{T}$, where

$$T_{ij} = \frac{1}{\rho} P \delta_{ij} - \tau_{ij}.$$

An intuitive feel for the role of the tensor $\boldsymbol{\tau}$ in the transport of momentum can be gained by looking closely at the case $i = 1$, $j = 2$. We relabel $u'_i = u'$, $u'_2 = v'$, $x_1 = x$, $x_2 = y$, and think of the coordinates $x_1$ and $x_2$ as being horizontal and vertical, respectively. Then the average $\langle u'v' \rangle$ appearing in (7) involves the horizontal and vertical components of the fluctuating part of the velocity. Imagine a particle whose instantaneous velocity has both a positive $x$-component $u'$ and a positive $y$ component $v'$. The former says that the particle is bearing some $x$-momentum, and the latter says that what it bears is being transported in the vertical ($y$) direction. The magnitude of this vertical transport of momentum is proportional to the product of the two components, $u'v'$. If one or both of these two components changes sign, this proportionality property is still seen to hold. Therefore $\langle u'v' \rangle$ is proportional to the mean vertical transport of $x$-momentum (which could be positive or negative) by means of turbulent fluctuations. This vertical transport represents the mean force per unit area exerted by the fluid above the point under consideration on the fluid below that point. Newton's first law characterizes a force as a rate of change of momentum; in this case, that rate of change is given by vertical transport of horizontal momentum-bearing particles. Finally, the $y$-derivative $\frac{\partial}{\partial y}\langle u'v' \rangle$, being a scaled difference of forces exerted at two nearby points, is the net $x$-component of the mean force per unit mass exerted on particles at that location by the turbulent fluctuations.

Generalizing these notions leads to an intuitive recognition of the last term in (8), $-\nabla \boldsymbol{\tau}$, as a force produced by turbulent fluctuations.

If we knew the tensor $\boldsymbol{\tau}$, then we could use that knowledge in (8) so that (8) and (5) would constitute a closed system for the determination of $\boldsymbol{U}$ and $P$. But nature is not so kind, and $\boldsymbol{\tau}$ is not known. Many attempts have been made in the past to remedy this basic defect, by writing down other equations for the determination of $\boldsymbol{\tau}$. These models propose a "closure" mechanism to produce a closed system for the mean values of various flow quantities. In all cases, these other equations are at least partly ad hoc, and in most cases partly empirical.

## 3  Early scaling concepts of Prandtl and von Karman

There have been many turbulence theories utilizing the ideas of Reynolds averaging; in fact the ones to be discussed in this paper are important examples. Prandtl and von Karman were, at an early stage, responsible for well known models in this category. Their concepts also included types of characteristic lengths, which are allied to the scaling theme to be developed below. Therefore we begin with brief discussions of some of their ideas.

In 1925, Prandtl [2] proposed a conceptual framework designed to provide insight into the mechanisms producing turbulence in shear flow, and also to provide relationships among key fluid dynamical quantities.

One aim of his was to relate the flux of momentum caused by turbulent fluctuations to the gradient of mean momentum. That there is such a relationship was simply an assumption springing from an analogy with conventional diffusion of quantities like heat in a stationary medium. In that setting, Fick's law asserts that the flux is proportional to the gradient of the concentration of the substance (like heat) that is being transported. In the present situation, since $\langle u'v' \rangle$ is (see previous section) a measure of the mean turbulent transport, in the vertical direction, of horizontal momentum, the analog of the heat equation for the mean $x$-momentum $U$ would be

$$\langle u'v' \rangle = -\epsilon \frac{dU}{dy}. \tag{9}$$

for some pseudo-diffusion coefficient $\epsilon$. That coefficient is generally called the *coefficient of eddy diffusion*, because one could argue that the transfer of momentum is caused by many eddies in the fluid.

To say more about $\epsilon$, Prandtl introduced the concept of *mixing length* $\ell$, which more or less represents the distance a fluid particle typically travels before transferring its momentum to another particle. Various versions of Prandl's theory have been involved with characterizing $\ell$ as it depends on local or global properties of the flow.

Since we will be dealing with vague definitions, the symbol "$\sim$" will be used to indicate a relation which is not precisely defined, but rather should be considered as part of a modeling concept. Moreover if numerical values are given to the two sides of the relation, they will be equal only in order of magnitude.

Prandtl's ideas related the left side of (9) directly to other local quantities as follows.

Suppose that the gradient $\frac{dU}{dy}$ is responsible, in a linear manner, for the local characteristic magnitude of the velocity fluctuations: $|u'| \sim k_1 \left|\frac{dU}{dy}\right|$; $|v'| \sim k_2 \left|\frac{dU}{dy}\right|$. The proportionality coefficients $k_i$ in these relations have to have the dimensions of length, and the most natural local characteristic length is the mixing length $\ell$. So we set $k_i \sim \ell$ to obtain

$$|u'| \sim \ell \left|\frac{dU}{dy}\right|; \quad |v'| \sim \ell \left|\frac{dU}{dy}\right|. \tag{10}$$

Then the average $|\langle u'v' \rangle|$ will be related to the quantity on the right of (10) squared:

$$|\langle u'v' \rangle| \sim \ell^2 \left|\frac{dU}{dy}\right|^2. \tag{11}$$

Now apply this to (9), at the same time ensuring that signs are chosen so that $\epsilon > 0$. We find that

$$\epsilon \sim \ell^2 \left|\frac{dU}{dy}\right|, \tag{12}$$

so that

$$\langle u'v' \rangle \sim -\ell^2 \left|\frac{dU}{dy}\right| \frac{dU}{dy}. \tag{13}$$

It seems intuitive that the local distance $\ell$ should grow shorter as one draws near the wall, because the wall exerts a constraint on the motions; and Prandtl sometimes proposed a linear relation

$$\ell \sim y. \tag{14}$$

This is probably not a good approximation very near the wall, because it would predict that $\langle u'v' \rangle$ grows like $y^2$ as one moves away from the wall, whereas there is abundant evidence that the growth rate is like $y^3$.

This, then, is how Prandtl and others proposed handling a truly complex phenomenon (forces caused by the turbulent transport of momentum) by relating it to a much simpler quantity (the gradient of mean velocity). This proposal had very limited success in explaining the processes of turbulence, although there was often good correlation with experimental data.

Von Karman [3] proposed his similarity hypothesis in 1930; a prominent ingredient was the assumption that $\epsilon$ and $\ell$ should be characterizable in terms of local properties of the fluid. For example, (14) would not qualify because the distance $y$ from the wall is not a local property. One might argue that the similarity hypothesis may approximate conditions far enough away from the walls and centerline.

This hypothesis proceeds from rescaling considerations, the idea being that if lengths for the velocity fluctuations in a neighborhood of a point in the flow are rescaled with a characteristic length $\ell$ (which we can identify as essentially the mixing length above), then key hydrodynamic quantities in that neighborhood are, except for an appropriate rescaling factor, functions only of the rescaled lengths; not of the position in the flow.

One then searches for a local quantity with the dimensions of length that one could use to characterize $\ell$. The possibly simplest choice would be the ratio

$$\ell \sim -\frac{dU}{dy} \bigg/ \frac{d^2U}{dy^2} \ . \tag{15}$$

(The minus sign comes about from noting that the ratio itself is always negative.)

Note that if this and (14) are both true, one could solve for $U(y)$ to get

$$U(y) \sim C_1 |\ln y| + C_2, \tag{16}$$

an example of the renowned logarithmic profile.

Other theories, as we shall see, are in partial agreement with this result and suggest other information as well. In particular, the theories brought out here in Secs. 5 and 6 involve ideas reminiscent of those of Prandtl and von Karman, but the differences outweigh the similarities.

The main approaches to wall-induced turbulence that we shall examine are heavily involved with various scaling concepts, and so we make a digression here to discuss some ideas basic to that subject.

## 4   The notion of scaling

Much of this section will consist of a review of scaling concepts, mostly well known, relating to the remainder of the paper. It will be useful to give a formal definition of scaling patch (less well known) in Sec. 4.1 and to formulate many of our results in terms of those patches. For example, it will be important to know that the overlapping regions of the classical example sketched in Secs. 4.2 and 4.3 are not scaling patches. The classical Izakson-Millikan observation covered in Sec. 4.6 is best known in the setting of turbulent flows, but is given here in a more general mathematical context and in the form of an approximative statement (Sec. 4.7) which is mathematically rigorous. The implications for wall-induced turbulent flow are discussed later in Sec. 5.6 and, more importantly, in Sec. 5.7.

Models involving small or large parameters are commonplace in the natural sciences; in more cases than not there are processes making up the action which operate on more than one, often many, different space and time scales. The phenomenon being studied can then most clearly and naturally be represented, in certain subdomains, in terms of functions of "rescaled variables", or of a combination of rescaled variables. Here rescaling means that new dependent and independent variables are defined, in differential form, as linear transformations of the original ones, the coefficients in the transformation generally being functions of the original small or large parameters.

Multiscaling refers to the event that more than one scaling are appropriately used, either in different subdomains or simultaneously in the same subdomain.

Our focus will be on differential equations containing parameters which are supposed to be small or large.

The ultimate goal here will be to gain some understanding of fluid motions by applying scaling concepts to the averaged Navier-Stokes equations. The recognition that the dynamics of turbulent fluids operate on a great many space and time scales has been a cornerstone of well known investigations into those processes, including the construction of mathematical models. Our goal is more limited, yet still daunting: to surmise information about important mean flow quantities on the basis of the averaged, rather than the original, Navier-Stokes equations. Rather than positing, at each specific location in the flow domain, an array of length scales, as would be the case for the microscopic turbulent motions, the mean flow profiles will themselves have unique characteristic lengths associated with each such location.

Our principal technique will be, first, to attempt to ascertain the local scaling behavior of those mean quantities by means of the averaged equations together with judicious assumptions, and then to derive further information about the mean profiles themselves.

In most of the following, there will be one independent small parameter called $\epsilon$, although another parameter $\beta$, which will also be small, will appear in some sections. The coefficients in the scaling transformations will depend on $\epsilon$, and possibly on $\beta$, and their orders of magnitude will be of primary importance. The notation $O(1)$ will refer to a quantity or function which generally depends on $\epsilon$, but is bounded above and below by positive constants independent of $\epsilon$ and $\beta$ as those parameters approach 0. When the lower bound is not assumed, we will generally write $\leq O(1)$. The meaning of other order relations hopefully will be clear from the context.

## 4.1 Scaling patches

We shall be dealing with functions of a single independent variable, and so the following discussion will apply to that case only. Moreover, the independent variable will represent a space coordinate. Finally, all statements about magnitudes in the following are to be understood in the order of magnitude sense relative to the small or large parameter under consideration. For example if $\epsilon$ is a small parameter, then $\epsilon$ has smaller order of magnitude than $\epsilon^{1/2}$ as $\epsilon \to 0$. To repeat what was said above, we shall use the convention that a quantity is $O(1)$ with respect to $\epsilon$ if it is bounded, and also bounded away from zero, by positive constants independent of $\epsilon$ as $\epsilon \to 0$.

Consider an interval in the domain of the independent space variable, and a single rescaling (of dependent and independent variables) in that subdomain. The interval may depend on the scaling, hence on $\epsilon$, and we assume the interval has size $O(1)$ when measured in the rescaled distance variable.

A "characteristic length" for the interval can be defined in terms of the scaling of the original space variable (call it $x$), which produces a new space variable (call it $\hat{x}$). Using differentials, we have, say, $dx = \alpha \, d\hat{x}$, with scaling coefficient $\alpha$. Then a characteristic length for that subdomain will be $\alpha$. As $\hat{x}$ traverses the $O(1)$ length of the interval, $x$ changes by an amount $O(\alpha)$. The stipulation that the size of the subdomain (called a "patch" below) be of size $O(1)$ in the rescaled variable is designed thereby so that $\alpha$ is a proper definition of characteristic length. Importantly, a patch cannot be artificially enlarged by adjoining a section in which the characteristic length is larger.

The rescaling (including that of the dependent as well as of the independent variables) can be thought of as being "natural" for a given problem if, when the solution is expressed in terms of the rescaled variables, the scaled dependent variables are seen to undergo, in that subdomain, variations which are not too large and not too small. This rate of variation could be gauged by the magnitude of the rescaled derivatives. The requirement "not too large" then would be taken to mean that all derivatives, of orders 1 up to some appropriate order, are bounded in magnitude independently of the parameters in the problem. It usually happens that some of these derivatives are necessarily zero or very small in places, so the corresponding (opposite) criterion cannot be imposed to gauge the satisfaction of the requirement "not too small". Instead, a proper criterion would be that the characteristic length $\alpha$ associated with the scaling under consideration cannot be decreased without the above criterion "not too large" being violated. In this, "decreased" means in the order of magnitude sense: if $\alpha$ is replaced by a different function of $\epsilon$ with smaller order of magnitude, so that in the newly scaled variables the derivatives are correspondingly larger, then the magnitudes of some of these derivatives must not be bounded independently of $\epsilon$.

Such an interval, together with its natural scaling, will in the following be called a *scaling patch*.

This appears to be a reasonable meaning for the concept of natural scaling in patches, but it ignores, so far, the important question, how does one find the scaling patches? That is, how does one determine the proper scaling in the proper locations? The search for scaling patches in wall-bounded turbulent flow will be the primary activity in Section 6. But it is not an easy question in general, and there are few easy mathematically rigorous criteria which can be applied to determine them. Nevertheless, there are nonrigorous arguments, most easily introduced through examples. Here is a straightforward one.

## 4.2 A classical example

The following is an elementary textbook model example of a problem in which scaling considerations, and in particular scaling patches, are very pertinent to understanding salient features of the solution. The solution, it turns out, can be written down explicitly, but we choose a different approach in order to better bring forth the ideas. Although it is well known, this example is chosen because we want to put forward a slightly nonstandard point of view, and it bears some similarity with the much more difficult averaged wall-induced turbulence problems which will be discussed later. Both scenarios have traditional inner and outer scaling regions. On the other hand there are also striking differences.

Let $\epsilon$ be a small positive parameter, i.e. $0 < \epsilon \ll 1$. We wish to solve the following boundary value problem for $u(\eta)$:

$$\epsilon^2 \frac{d^2 u}{d\eta^2} - u + g(\eta) = 0 \quad \text{for} \ \ 0 < \eta < 1, \tag{17}$$

$$u(0) = \alpha, \quad u(1) = \beta, \tag{18}$$

where $\alpha$ and $\beta$ are fixed numbers and $g$ is any given smooth function. To make things simple, we suppose that $g$ is independent of $\epsilon$, and that it is not a constant (i.e. it has nontrivial variation). The solution of course will depend on $\epsilon$ as well as on $\eta$, and the desire is to find that dependence for <u>all</u> small enough values of $\epsilon$, say for $0 < \epsilon \leq \epsilon_0$, where $\epsilon_0$ is a fixed small number. This dependence of the solution on both $\eta$ and $\epsilon$ is sometimes expressed by writing $u(\eta; \epsilon)$.

As a first guess for an approximate solution, we try discarding the first term in (17), which has a small factor $\epsilon^2$. We obtain the "reduced" problem

$$u = g(\eta). \tag{19}$$

The function $u = g(\eta)$ approximately satisfies (17) in a formal sense, but is generally far from satisfying the two boundary conditions (18) (unless by unlikely accident $g(0) = \alpha$ or $g(1) = \beta$). So we must either scrap that solution altogether, or doctor it up. The latter is possible.

This is where the search for scaling patches comes in. Two things are clear: the locations of the trouble are at the two boundaries $\eta = 0$ and $\eta = 1$. And secondly, somehow the discarded term $\epsilon^2 \frac{d^2u}{d\eta^2}$ at those two locations must be important after all. We try constructing an "internal" patch by excluding neighborhoods of the two troublesome boundary points as follows. Take any number $\delta > 0$ independent of $\epsilon$, and use, for the patch, the original scaling, leading to (19), in the subdomain defined by $\{\eta : \delta < \eta < 1 - \delta\}$. This interval, together with the original scaling in which the variables $\eta$ and $u$ remain unchanged, will be the first scaling patch. To lowest approximation in the parameter $\epsilon$, the differential equation (17) becomes (19). An important part of our argument later will be centered around the issue of how far we can enlarge this interval by letting $\delta$ depend on $\epsilon$.

Now let us also try to construct a scaling patch which encompasses the left boundary $\eta = 0$. The appearance of the differential equation can be changed by rescaling in such a way as to render the derivative term overtly important. The most natural way is by passing to a new independent variable $y$ by

$$y = \frac{\eta}{\epsilon}, \quad U = u. \tag{20}$$

Then we consider $U$ to be a function of $y$ in some subdomain consisting of values of $\eta$ near 0. We shall be conservative at first and take that subdomain as $\{0 \leq y < y_0\}$, where $y_0$ is independent of $\epsilon$. Later, we see about extending the interval by letting $y_0$ depend on $\epsilon$

The rescaled differential equation is

$$\frac{d^2U}{dy^2} - U(y) + g(\epsilon y) = 0. \tag{21}$$

Neglecting $\epsilon$ in (21) leads to

$$\frac{d^2U}{dy^2} - U(y) + g(0) = 0. \tag{22}$$

The boundary condition at $\eta = 0$ (the same as $y = 0$) now becomes

$$U(0) = \alpha. \tag{23}$$

It is proposed, then, that the scaling given by (20) in the interval $\{0 \leq y < y_0\}$ be our second scaling patch, and that the solution in that patch be approximately a solution of (22) and (23). Moreover it is reasonable to impose the condition that $U$ be bounded when $y$ grows large (like $O(1/\epsilon)$). This provides a unique solution and the formal approximation

$$U(y) \approx g(0)\left(1 - e^{-y}\right) + \alpha e^{-y}. \tag{24}$$

This, of course, is only an approximation which is put forward for further verification. For one thing, notice that $\lim_{y \to \infty} U(y) = g(0)$ and that the original approximation $u = g(\eta)$ is also close to the value $g(0)$ for $\eta \ll 1$. Thus in a sense the approximation at the left boundary meshes smoothly with the supposed approximation in the interior of the interval. Let us dwell on the idea of smooth meshing a little more carefully. Let $\omega$ be any positive number, arbitrarily small. We will show that there are large values of $y$ and small values of $\epsilon$ for which the boundary approximation $U(y)$ and the interior approximation $g(\eta) = g(\epsilon y)$ are both closer than $\omega$ to $g(0)$, hence arbitrarily close to each other. Let $y_1$ be a large number, depending on $\omega$, such that $|U(y) - g(0)| < \frac{1}{2}\omega$ for all $y > y_1$. Such a number exists, as you can see from (24). Next, let $\eta_1$ be a small number, again depending on $\omega$, such that $|g(\eta) - g(0)| < \frac{1}{2}\omega$ for all $\eta < \eta_1$. Such a number $\eta_1$ exists because of the continuity of the function $g$. Since $\eta = \epsilon y$, both of these statements will be valid if $\epsilon$ is chosen such that $\epsilon < \frac{\eta_1}{y_1}$, and $y$ is in the interval $y_1 < y < \eta_1/\epsilon$. Since both are valid for these values of $y$, necessarily $|U(y) - g(0)| < \omega$. The conclusion is that if $\epsilon$ is small enough, depending on $\omega$, there are places where the inner and outer approximations are closer together than $\omega$, which can be chosen as small as desired. This is what we mean by the two approximations meshing smoothly together.

In summary, we have found (a) what appears to be a reasonable approximation for the solution $u(\eta; \epsilon)$ in regions of the interval $[0, 1]$ not too close to either boundary, and also (b) a reasonable approximation in regions close to the left boundary.

The same procedure yields a third scaling patch near the right hand boundary. The solution in it undergoes a similar exponential (in terms of a rescaled variable) transition from the value $\beta$ to $g(1)$ as we move left from that point. The rescaling in that case is

$$z = \frac{1 - \eta}{\epsilon}, \quad V(z) = u(\eta) = u(1 - \epsilon z),$$

and the right inner solution in that patch is

$$V(z) \approx g(1)\left(1 - e^{-z}\right) + \beta e^{-z}. \tag{25}$$

In the interior, the solution is approximately a function of only $\eta$ with no $\epsilon$-dependence, so no new scaling is necessary. In a region which excludes small neighborhoods of each boundary (we will discuss how small later), the original variables are the natural ones in our adopted sense. That region, together with the original unscaled variables, constitutes one scaling patch. The small interval $y < y_0$, i.e. $\eta < \epsilon y_0$), together with the rescaling (20), constitutes a second scaling patch, in which the solution undergoes, in an exponential manner, a transition from the imposed value $\alpha$ to the value $g(0)$ associated with the first scaling patch, but extended to the forbidden boundary point. At this point the two patches do not touch each other but, as we shall see, the ranges of validity of the corresponding approximations can be enlarged so that they overlap.

Thus through proper "scaling", the original problem may be clarified and simplified in certain subranges of the range of independent variables.

The outlined procedure for finding scaling patches in the example problem is only heuristic, but it can be made rigorous in this and other cases. Moreover, we have only found the most basic (lowest order) approximation. Successive higher order approximations, with errors of orders $O(\epsilon)$, $O(\epsilon^2)$, etc., can be found and proved to be correct in this and in a large class of similar problems, such as elliptic boundary value problems when there are several independent variables.

## 4.3 Inner, outer, and overlapping regions

We have constructed independent approximate representations of the solution in three subdomains: (19) for $\eta$ not too close to either boundary point, (24) near the left boundary, and (25) near the right one.

Traditionally, these three approximate solutions are called the *outer, left inner, and right inner* solutions. We denote the outer solution by $u_o(\eta)$ and the left inner by $U(y)$. In the following, we shall usually disregard the right inner approximation because its analysis follows that of the left inner solution.

We now ask about the possibility of extending those subdomains, while retaining the validity of the approximate representations. We do this by introducing a family of possible intermediate scalings parameterized by an exponent $\gamma$ in the range $0 < \gamma < 1$. The independent variable $\eta$ is rescaled to obtain a new variable $s$ by

$$\eta = \epsilon^\gamma s, \quad y = \epsilon^{\gamma-1} s. \tag{26}$$

We are excluding the choice $\gamma = 0$, which would correspond to the original scaling, and $\gamma = 1$, corresponding to (20). Set $U_\gamma(s) = u(\eta) = u(\epsilon^\gamma s)$.

The rescaled differential equation (17) becomes

$$\epsilon^{2-2\gamma} \frac{d^2 U_\gamma}{ds^2} - U_\gamma + g(\epsilon^\gamma s) = 0, \tag{27}$$

and the boundary condition is $U(0) = \alpha$. We ask, can the region defined by $\{0 < s_0 < s < s_1\}$, where $s_0$, $s_1$ are independent of $\epsilon$, be the location of a new scaling patch? (The answer will turn out to be no in the strict sense, but we can try.) If so, the corresponding solution should, to lowest order as $\epsilon \to 0$, hence $\epsilon^\gamma \to 0$, satisfy (27), which to lowest order comes out to be

$$U_\gamma(s) = g(0) \tag{28}$$

(since $0 < \gamma < 1$). That is a very simple approximate solution indeed.

Keeping that in mind, we now examine the left inner and the outer solutions (24) and (19) in the proposed patch. For the left inner, we obtain the expression $U\left(\epsilon^{\gamma-1} s\right) \approx [g(0)(1 - e^{-y}) + \alpha e^{-y} g(0)]_{y=\epsilon^{\gamma-1} s}$. To find the lowest order version of this expression, just let $\epsilon \to 0$. We get, again to lowest order, the same result as (28):

$$U\left(\epsilon^{\gamma-1} s\right) \approx g(0). \tag{29}$$

In the case of the outer solution, a similar procedure yields

$$u_o\left(\epsilon^\gamma s\right) \approx g(0). \tag{30}$$

This means that in any of the *proposed* intermediate scaling patches (as it will turn out, they are not true patches), the inner and outer solutions are both approximate solutions of the rescaled equation (27), so that in this sense each of them is a valid approximation to the true solution, for small enough $\epsilon$.

This verifies that the original subdomains of the inner and outer scalings can be extended to include the intermediate regions, the corresponding approximations continuing to be valid. In other words, there is a region of overlap, in which both the inner and outer solutions are valid. In fact, there are many overlap regions, depending on the choices of $\gamma$, $s_0$, and $s_1$. These are all approximate solutions, and the main effective differences among them lies in the accuracy of the approximation. Bear in mind that $\epsilon^\gamma$ is approximated by 0 better when $\gamma$ is larger.

Finally, note that the approximate solutions in any of these overlap regions are not very interesting: they are all constant to lowest order, equal to $g(0)$! This implies that the regions $\{s_0 < s < s_1\}$, together with the scaling (26), do not form legitimate scaling patches! The reason is that the solution does not satisfy one of the stated criteria in section 4.1, namely that the length scale $\alpha$, which by (26) is equal to $\epsilon^\gamma$, can certainly be reduced without the scaled dependent variable, which we have seen is constant, having unbounded derivatives.

The same procedure yields overlapping regions near the right boundary point, in which the approximate solutions are all another constant, $g(1)$.

## 4.4 A uniform approximation

The separate approximations in the three specified regions, namely $u_o(\eta) = g(\eta)$ (19), $U(y)$ (24) and $V(z)$ (25), can be combined into a single expression, providing a uniform approximation throughout the interval $[0, 1]$. For this, we define

$$F(y) = U(y) - g(0), \quad H(z) = V(z) - g(1). \tag{31}$$

The uniform approximation is then

$$U_{unif}(\eta; \epsilon) = u_o(\eta) + F(y) + H(z) = u_o(\eta) + F(\eta/\epsilon) + H((1-\eta)/\epsilon). \tag{32}$$

Its validity can be verified directly in each of the regions considered above. This form consists of a sum of terms, each a function of one of the scaled variables.

## 4.5 A generalization

It is natural to ask whether the outlined features in the overlap region of that classical example apply also to a wider class of functions, and indeed they do. Rather than starting with a differential equation, we take a general class of functions expressed in the form of a sum of individual functions of the three scaled variables $\eta$, $y$, $z$ separately, as in (32). So we consider any function in the form (32), irrespective of whether it is a solution or approximate solution of some problem. Such a function may still have, depending on $u_o$, $F$ and $H$, an outer approximation and two inner approximations. It may also have a region of overlap between the outer and one of the inner solutions.

No generality is lost by taking $H = 0$, and we do that. Let $\epsilon \ll 1$, $y = \frac{\eta}{\epsilon}$, and $u_o(\eta)$ and $F(y)$ be any functions such that $u_o(\eta)$ has a limit as $\eta \to 0$ and $F$ has a limit as $y \to \infty$:

$$\lim_{\eta \to 0} u_o(\eta) = u_o(0); \quad \lim_{y \to \infty} F(y) = G. \tag{33}$$

We call $\eta$ the outer variable and $y$ the inner variable. Finally, let

$$u(\eta; \epsilon) = u_o(\eta) + F(y) = u_o(\eta) + F(\eta/\epsilon). \tag{34}$$

This will be the central prototypical two-scaled function which will be approximated differently in an inner and an outer region.

In a region $\{\eta > \delta\}$ for any fixed $\delta > 0$, $\eta/\epsilon \to \infty$ uniformly as $\epsilon \to 0$, so that $F$ can be approximated by $G$:

$$u(\eta, \epsilon) \approx u_o(\eta) + G \equiv u_{out}(\eta), \tag{35}$$

which we call the outer representation of $u$. Similarly near $\eta = 0$, we replace $\eta$ by 0 in the first term on the right of (34) to obtain the inner representation

$$u_{in}(y) \equiv u_o(0) + F(y). \tag{36}$$

Now consider a family of intermediate regions parameterized by $\gamma \in (0, 1)$ in which $s$, defined by (26), is confined to the interval $s_0 \leq s \leq s_1$, the $s_i$ being specified constants. In such a region, we find the approximations

$$u_{in} \approx u_o(0) + F(\epsilon^{\gamma-1} s) \approx u_o(0) + G, \tag{37}$$

$$u_{out} \approx u_o(\epsilon^\gamma s) + G \approx u_o(0) + G, \tag{38}$$

$$u(\eta, \epsilon) \approx u_o(\epsilon^\gamma s) + F(\epsilon^{\gamma-1} s) \approx u_o(0) + G. \tag{39}$$

The conclusions are, first, that the inner and outer representations are both valid in the intermediate region, which is therefore an overlap zone. Secondly, in the overlap region, the approximation is equal to the constant $u_o(0) + G$.

These conclusions are the same, in our generic class of examples, as those for the classical example in section 4.2.

## 4.6 The Izakson-Millikan observation

In [4] and [5], Izakson and Millikan independently considered functions of the form (34) without imposing limit conditions such as (33). Their context was the averaged equations of wall-induced turbulence, but their reasoning is valid in the present more general scenario. They showed, more or less, that an assumption of the existence of an overlap region is sufficient to imply that the approximations in that region are either constant or logarithmic. Beginning in Sec. 6, a different approach, but with a similar objective, will be discussed in detail. Its method, limited to the turbulence problem, will be based on a search for scaling patches

In the Izakson-Millikan argument, then, no special conditions are required at either boundary of the interval $I$ (below) on which our functions are defined. Moreover, instead of providing functions of inner and outer variables and asking whether there is an overlap region of validity of the corresponding inner and outer approximations, we now do just the opposite. We assume that there is such a region of common validity, and ask what more, if anything, that implies about those approximations. The answer is surprising.

Assume, for some unknown functions $u_o(\eta)$, $U(y)$ (where $y = \eta/\epsilon$ as always), $G(\epsilon)$ and interval $I$, that

$$u_o(\eta) + G(\epsilon) = U(y) \tag{40}$$

for all values of $\eta \in I$ and all values of $\epsilon$ in some interval $K$. Since $\eta$ and $\epsilon$ can vary independently of each other in their respective domains, $\eta$ and $y$ also vary independently of each other, for $\eta \in I$ and $y \in J$, $J$ here being the set of all values of $y = \eta/\epsilon$ for $\eta \in I$ and $\epsilon$ in its previously allocated interval of variation $K$. The assumed identity (40) asserts that there is a region of overlap where the outer function $u_o(\eta)$ (plus a correction $G(\epsilon)$) equals the inner function $U(y)$. All these functions are at present unknown. It turns out that the correction $G$ is necessary to include, because without it, (40) is too stringent a condition to be fulfilled in general.

First, in (40) make $\epsilon$ a fixed number in $K$, and of course $y = \eta/\epsilon$ for all $\eta \in I$. Differentiating (40) with respect to $\eta$ and multiplying by $\eta$, we obtain

$$\eta u_o'(\eta) = \frac{\eta}{\epsilon} U'(y) = y U'(y). \tag{41}$$

This is valid for every chosen value of $\epsilon \in K$, so is true for all $y \in J$ as well as $\eta \in I$. But since, as remarked before, $\eta$ and $y$ vary independently of each other, each side of (41) must be a constant. For example $\eta$ can be allowed to vary throughout $I$ while $y$ is held constant, which of course implies that the right side of (41) is held constant. That means the left side is also constant. Therefore for some constant $A$,

$$\eta u_o'(\eta) = A; \quad y U'(y) = A. \tag{42}$$

Integrating these two equations, we find, for some other constants $B$ and $C$,

$$u_o(\eta) = A \ln \eta + B, \quad U(y) = A \ln y + C = A \ln \eta + C - A \ln \epsilon, \tag{43}$$

the last equation a result of the identity $y = \eta/\epsilon$. Since $u_o$ is assumed to be a function only of $\eta$, necessarily $B$ is a constant independent of $\epsilon$, and the same holds true of C for a similar reason. The function $G$ can now be determined if we substitute the expressions (43) back into (40). Doing so while noting that $\ln y = \ln \eta - \ln \epsilon$ yields

$$G(\epsilon) = -A \ln \epsilon + C - B. \tag{44}$$

In short, the three functions in question must be of the form (43), (44), where $A$, $B$, $C$ are constants. The original assumed identity (40) dictates a great amount of information about these functions, while not specifying them exactly.

What this result says is that functions which are simultaneously regular functions of an inner and an outer variable, for all values of those variables in certain intervals, are either constant or logarithmic. If it is known that the function is not constant, then this property of simultaneity is equivalent to being logarithmic.

### 4.7 A more realistic version

The result of Izakson and Millikan is mathematically rigorous, in the sense that if (40) holds in some intervals as specified, i.e. if some function can be expressed exactly in terms of either an inner or an outer variable in some overlap region, then the logarithmic properties (43) and (44) also hold in that same region.

However, the assumption (40) probably never holds exactly in practice, so that the conclusion is vacuous. In any concrete situation, it will be far more reasonable to assume that (40) is only approximately satisfied. Then the suggestion, which is not yet proved, is that the functions involved will be approximately logarithmic.

Fortunately, such a conclusion can also be rigorously established, provided that appropriate senses are given to the two approximations (the one in the hypothesis, and the one in the conclusion). That will be done in this section. Note also that Gill in [6] provided such a proof under more involved assumptions and with entirely different methods.

In place of (40), we now assume that (40) holds with an additional small error term $r(\eta, y, \epsilon)$, which will be written $r(\eta, y)$, since $\epsilon$ can be expressed in terms of $\eta$ and $y$. The function $r$ is not known exactly, but that does not prevent estimates about the accuracy of (43) and (44) being deduced in terms of the magnitude of $r$ (and, as it turns out, of its derivatives). Thus we start with

$$u_o(\eta) + G(\epsilon) = U(y) + r(\eta, y). \tag{45}$$

As we shall show, although (43) and (44) no longer hold, one can still find upper and lower bounds of logarithmic type for the functions $u_o$, $U$, $G$ in terms of bounds on $r$ and its derivatives.

Differentiating (45) as before, one obtains

$$\eta u_o'(\eta) = y U'(y) + R(\eta, y), \tag{46}$$

where $R(\eta, y) = \eta r_\eta + y r_y$, subscripts denoting derivatives. Let $\rho$ be an upper bound for $|R|$, valid for all $\eta$, $y$ in the intervals $I$, $J$ respectively. Choose any $y_0 \in J$ and let $A = y_0 U'(y_0)$. Then setting $y = y_0$ in (46), we get

$$|\eta u_o'(\eta) - A| \leq \rho. \tag{47}$$

This holds for all $\eta \in I$. Also from (46),

$$y U'(y) - A = \eta u_o' - A - R,$$

so that by (47)
$$|yU'(y) - A| \leq |\eta u'_o - A| + \rho \leq 2\rho. \tag{48}$$

This says that when $\rho$ is small, the quantities on the left sides of (42) are almost constant, the deviation from constancy being no greater than $\rho$ and $2\rho$, respectively. Dividing (47) by $\eta$ and rearranging terms, we get

$$\frac{A - \rho}{\eta} \leq u'_o(\eta) \leq \frac{A + \rho}{\eta},$$

and after integrating,
$$(A - \rho) \ln \eta + B_1 \leq u_o(\eta) \leq (A + \rho) \ln \eta + B_2, \tag{49}$$

where $B_i$ are integration constants. In this sense, $u_o$ is almost logarithmic when $\rho$ is small. In the same way we obtain
$$(A - 2\rho) \ln y + C_1 \leq U(y) \leq (A + 2\rho) \ln y + C_2, \tag{50}$$

with similar upper and lower bounds for $G$.

In short, assuming $R$, i.e. $\rho$, is small in the intervals selected, one concludes that $U(y)$ is bounded above and below by logarithmic expressions with coefficients of the logarithm functions which are close to each other, the discrepancy being smaller than $2\rho$. It must be admitted, however, that one generally does not know the nature of the error term $R$, its magnitude $\rho$, or indeed the interval of overlap, if any. If the intervals $I$, $J$, hence $K$ are chosen to be shorter and well placed, one might expect $\rho$ to be smaller and therefore the logarithmic approximations to be better. There is a tradeoff between the accuracy of the logarithmic approximation and the size of the interval where that approximation is valid. It can be guessed that as one moves closer to the canonical outer domain, $r$ increases in magnitude because although the outer approximation $u_o(\eta) + G(\epsilon)$ is more exact, the inner expression $U(y)$ is not a good approximation, so the two cannot be close to each other, as (45) would seem to imply. Somewhere between the outer and the inner domains, $r$ would be minimal, but still not zero. This was with $\epsilon$ fixed. A natural further question, therefore, is whether, in such fixed intervals of the space variable $\eta$, the error $r$ and its relative $R$ approach 0 as $\epsilon \to 0$. The present argument does not answer that question, but an argument in support of a similar conclusion in the context of wall-induced turbulence will be given below in Sec. 6.6.

## 5 The mean structure of turbulent Couette flow in a channel

At this point we leave the mathematical digression about scaling and turn to the main issue of this paper, namely the application of scaling ideas to turbulence induced by wall friction. Turbulent Couette flow is possibly the simplest nontrivial example. From this point through Sec. 5.6, we will be repeating well-known standard arguments but with different emphases. Newer, hence lesser known, material will be found in sections beyond that.

Denote the components of $\boldsymbol{x}$ and $\boldsymbol{u}$ by $(x, y, z)$ and $(u, v, w)$ respectively. Consider a channel bounded on top and bottom by horizontal planes $\{y = 0; \ y = 2\delta\}$. The top plane moves with given steady velocity in the streamwise direction $x$, while the lower plane remains stationary. This causes a shear stress in the fluid between the two planes. In the mean, fluid particles which are vertically aligned at one moment of time slide past each other horizontally.

If the velocity is sufficiently large, the resulting shear causes the flow to be turbulent. We seek to understand the "scaling structure" of the mean velocity of the fluid, and other quantities, when the flow has reached "equilibrium". The scaling analysis of turbulent Couette flow described here, was developed in [7, 8, 9, 10, 11]. Here <u>scaling structure</u> will mean that we will look for scaling patches for the mean velocity and Reynolds stress profiles, or other relevant ways to scale those functions. And the term equilibrium will refer to the state in which the mean velocity is everywhere horizontal and depends, with Reynolds stress, only on the normal coordinate $y$. Although there will be perhaps violent particle fluctuations in all directions, the prevailing (mean) motion will be only horizontal.

The dependence only on $y$ implies that all averaged quantities are uniform along the channel and do not change in time. The upshot is that the averaged Navier-Stokes equations have only one independent variable ($y$).

Before we get into the turbulence analysis, consider the corresponding laminar flow, in which all fluctuation parts vanish, and in fact the velocity has only an $x$-component, which is simply a linear function of $y$, the coefficients being adjusted to match the given velocities of the bounding planes. This is a very simple solution, and can be verified to be a solution of the Navier-Stokes equations (1) with 0 pressure gradient. In fact the inertia terms of the conservation equation for the $x$-component of momentum vanish automatically because each velocity component is either 0 or has $x$-derivative 0. For a similar reason, the viscosity terms also vanish.

The analogous problem for turbulent flow is orders of magnitude more difficult, and can only be solved imprecisely. We will examine it in the framework of Reynolds averaging.

But a possible paradox appears. It was just brought out that the eminently simple laminar flow, in which $u$ is a linear function of $y$, is an exact solution of the Navier-Stokes equations, and those equations govern the flow, laminar or turbulent. So if we have a solution, why do we need to look for a "turbulent" one? The answer is because (a) as we set up the problem, there are more solutions than just the laminar one; solutions are not unique; and (b) the laminar one happens to be very unstable at high Reynolds numbers, therefore not seen in practice and hence unphysical.

## 5.1 The differential equations

To proceed, suppose the turbulent flow to be "fully developed," statistically stationary and two-dimensional in the channel. All averaged quantities except pressure do not depend on $x$, $z$, or $t$; just on $y$. Moreover, we suppose that the only nonzero component of the mean velocity will be the $x$-component, which we denote by $U(y)$. (This part is the same as the laminar case.)

The $x$ and $y$ components of (8) are greatly simplified, because of the independence of $\boldsymbol{\tau}$ on $x$, $z$, $t$:

$$-\nu \frac{d^2 U}{dy^2} + \frac{1}{\rho}\frac{\partial P}{\partial x} + \frac{d}{dy}\langle u'v'\rangle = 0, \tag{51}$$

$$\frac{1}{\rho}\frac{\partial P}{\partial y} - \frac{d}{dy}\langle (v')^2\rangle = 0. \tag{52}$$

At the two walls $\{y = 0, \ y = 2\delta\}$, all the fluctuating parts of $\boldsymbol{u}$ are 0, so the Reynolds stress $\langle u'v'\rangle$ vanishes as well. At the lower wall, $U = 0$ and its derivative is related to the frictional stress exerted on the wall by the fluid (or vice versa).

We have here two equations for the four unknown functions $U$, $P$, $\langle u'v'\rangle$ and $\langle (v')^2\rangle$. Actually, Couette flow is characterized by the absence of an applied pressure gradient; the sole impetus for the flow is the differential motion of the two walls. Therefore we set $\frac{\partial P}{\partial x} = 0$ ($P$ itself depends on $y$, however, as one can see from (52)). In the case of other steady channel flows, $\frac{\partial P}{\partial x}$ may not be zero, but it would turn out necessarily to be constant.

The equation (51) becomes

$$-\nu \frac{d^2 U}{dy^2} + \frac{d}{dy}\langle u'v'\rangle = 0. \tag{53}$$

It is important to recognize that this is a simple balance of forces, which must occur at each point of this steady (in the mean) flow. The two forces exerted on the fluid are (1) the friction, or viscous, force $\nu \frac{d^2 U}{dy^2}$, and (2) the force $-\frac{d}{dy}\langle u'v'\rangle$ caused by the turbulent fluctuations, i.e. the gradient of the appropriate Reynolds stress, by which $x$-momentum is transported in the $y$ (perpendicular) direction by the fluid's fluctuations in both directions. An intuitive description of the two forces can be given. (1) The derivative $\frac{dU}{dy}$ measures the magnitude of the shearing motion, i.e. the rate at which nearby $x$-directed particles are moving relative to one another. This shearing magnitude, times the kinematic viscosity $\nu$, is proportional to the force per unit area that the fluid lying above the horizontal plane through the point under consideration is exerting on the fluid below (and vice versa). This is called the shear stress. The difference between the shear stress at two nearby values of $y$, say $y_1$ and $y_2$, is the net effective force per unit area experienced by the slab of fluid in the intermediate region $\{y_1 \leq y \leq y_2\}$. Dividing this expression by $y_2 - y_1$ and passing to the limit, we obtain the net force per unit mass acting on fluid particles at that location, and see that it equals $\nu \frac{d^2 U}{dy^2}$. Again, this is termed the viscous force.

(2) As shown in Sec. 2, the $y$-derivative $\frac{d}{dy}\langle u'v'\rangle$ is the $x$-component of the mean force per unit mass exerted on particles at that location by the turbulent fluctuations.

Since there are only these two forces in balance, it can be said, contrary to some assertions, that the viscous forces and those due to Reynolds stresses are both equal players, hence both important, everywhere in the flow.

At this point we have reduced the problem to a single equation (53) for $U$ and $\langle u'v'\rangle$. There are still more unknowns (2) than equations (1). But we shall nevertheless be able to at least surmise some important information just from the simple mean balance law (53) plus known boundary conditions, in concert with educated assumptions.

## 5.2 The friction velocity and boundary conditions

We start with some crucially important concepts and constants associated with the interaction of the fluid with the wall. Let $\tau_w$ denote the mean stress exerted on the wall by the fluid flowing past it. It is proportional to the viscosity $\mu$, as well as the magnitude of the mean velocity's shear at the wall. It is given by

$$\tau_w = \mu \frac{d}{dy}\langle u\rangle = \mu\left[\frac{dU}{dy} + \frac{d}{dy}\langle u'\rangle\right] = \mu\frac{dU}{dy}, \qquad (54)$$

since $\langle \frac{d}{dy} u'\rangle = \frac{d}{dy}\langle u'\rangle = 0$. At $y = 0$, $\nu \frac{dU}{dy} = \frac{1}{\rho}(\mu \frac{dU}{dy}) = \frac{1}{\rho}\tau_w > 0$. Now the quantity $\frac{1}{\rho}\tau_w$ has the dimensions of velocity squared; it therefore defines a characteristic velocity $u^*$, called the *friction velocity*, by

$$\frac{1}{\rho}\tau_w = (u^*)^2. \qquad (55)$$

From this relation and the previous one, we can express

$$\frac{dU}{dy} = \frac{u^{*2}}{\nu} \quad \text{at } y = 0. \qquad (56)$$

This, together with the stipulation that

$$U(0) = 0, \qquad (57)$$

provide a pair of boundary conditions at the wall for $U(y)$. It seems natural to treat the velocity of the upper wall as a given quantity to be built into the mathematical formulation; but analytically the simpler route is instead to think of the velocity $u^*$ (in (56)) as given, and the upper wall velocity as to be determined. It is clear from physical considerations that either of these two velocities is a monotone function of the other.

As noted before, the fact that all fluctuations vanish at the wall imply similar conditions for the Reynolds stress. In fact,

$$\langle u'v' \rangle = \frac{d}{dy} \langle u'v' \rangle = \frac{d^2}{dy^2} \langle u'v' \rangle = 0 \tag{58}$$

at $y = 0$. The first two conditions here follow from $u' = v' = 0$ at the wall. The third condition comes about because, using subscripts $y$ to denote derivatives, we have $\langle u'v' \rangle_{yy} = \langle u'_{yy} v' \rangle + 2 \langle u'_y v'_y \rangle + \langle u' v'_{yy} \rangle$. The first and third terms in this expression vanish at the wall because $u'$ and $v'$ do; and the second term also vanishes because (6) together with $u'_x = 0$ implies $v'_y = 0$.

Finally at the center of the channel at $y = \delta$, we invoke the symmetry of the flow to conclude that

$$\frac{d}{dy} \langle u'v' \rangle = 0 \quad \text{at } y = \delta, \tag{59}$$

since every vertical fluctuation $v'$ is matched, and canceled during the averaging process, by another one in the opposite direction. From (53), $U$ has an inflection point. In fact, consistent with (59), the Reynolds stress $\langle u'v' \rangle$ increases in magnitude from 0 at the wall to a maximal value at the centerline.

## 5.3 Dimensionless variables in the core region

For better insight, we now seek to nondimensionalize (53). That is, we rescale the variables by multiplying them by typical and meaningful characteristic dimensional constants so that the results are dimensionless. There are at least two traditional and natural ways to do this. In any case, we argue that the characteristic velocity in this turbulent flow should probably be $u^*$, because the former's magnitude should be directly related to the wall stress, which is what slows the fluid down at the lower wall and causes it to go forward at the upper wall, and so in some sense causes the turbulence. So we nondimensionalize $U$ by $u^*$. (As brought out above, another natural, but less convenient velocity unit would be the maximum mean velocity or the average mean velocity.)

In the interior of the channel, it is intuitive that $y$ should probably be scaled by the channel half-width $\delta$, simply because that is the only immediately obvious characteristic length appropriate to that part of the channel (further justification will be given in Sec. 6.8). And everywhere, the Reynolds shear stress should be scaled by the shear stress at the boundary, so by $(u^*)^2$. We therefore define

$$U^+ = \frac{U}{u^*}, \quad \eta = \frac{y}{\delta}, \quad T = -\frac{\langle u'v' \rangle}{(u^*)^2}. \tag{60}$$

The centerline of the flow region is at $\eta = 1$, and by symmetry considerations, the mathematical problem can be set up in the half-channel $0 \leq \eta \leq 1$. The superscript "+" in $U^+$ is the traditional way to signify that this variable is normalized with parameters related to what happens at the wall: the friction velocity $u^*$ and/or $\tau_w$. The scaled distance $\eta$ is the first natural way to nondimensionalize the $y$-coordinate, and $\eta$ will be called the "outer distance variable". This, along with the "inner distance variable" $y^+$ to be discussed in Sec. 5.4 below, have been standard choices for dimensionless distance since the beginning of theoretical investigations of wall-induced turbulence. However, the existence of scaling patches as defined in Sec. 4.1 employing these scaled distances remains to be established. That will be done in Section 6.8.

With the normalization (60) we obtain in place of (53),

$$\frac{dT}{d\eta} + \frac{\nu}{u^*\delta}\frac{d^2U^+}{d\eta^2} = 0. \tag{61}$$

We also use $u^*$ and $\delta$ to define our Reynolds number $R^* = \frac{u^*\delta}{\nu}$ and small parameter $\epsilon^2 = (R^*)^{-1}$. Using this notation, we get

$$\frac{dT}{d\eta} + (R^*)^{-1}\frac{d^2U^+}{d\eta^2} = \frac{dT}{d\eta} + \epsilon^2\frac{d^2U^+}{d\eta^2} = 0. \tag{62}$$

As noted earlier, we have an underdetermined problem—a single equation for the two unknowns $T$ and $U^+$. But there is even more bad news: the boundary condition at $\eta = 0$ is in trouble. Naively assuming the formally small second term can be neglected across the whole channel, we get that $T$ is constant, and since it vanishes at the wall, we would obtain that $T = 0$ everywhere. This is incorrect, of course, for a reason similar to that which applies to the classical example in Sec. 4.2: a different scaling applies near the wall.

We construct now a second scaling domain, near the wall, to partially remedy this defect, as was done for the classical example in Sec. 4.2.

## 5.4 The wall layer, rescaling, and law of the wall

It is traditionally recognized that we do have at least two space scales in the channel—one of them associated with $\delta$ and another one close to the wall; we'll get to that shortly. We will have an "outer" and an "inner" approximate solution. Finding out how to connect them makes an interesting story. However, we cannot carry out a full-blown asymptotic analysis because too much is unknown. Much of the following proceeds by "reasonable suppositions", which can also be verified to some extent by experiments.

It is natural to choose the inner scaling in such a way that the two terms on the left of (62) have the same orders of magnitude. After all, those two terms represent the two forces in the fluid which have to balance (note, that's not what we did to get (61)). Thus we define

$$y^+ = \eta R^* = \frac{u^*}{\nu}y. \tag{63}$$

Then (62) becomes

$$\frac{dT}{dy^+} + \frac{d^2U^+}{dy^{+2}} = 0. \tag{64}$$

In this, there is no incompatibility with the boundary conditions at $y^+ = 0$:

$$T = 0 \text{ and } \frac{dU^+}{dy^+} = 1 \text{ at } y^+ = 0. \tag{65}$$

The integrated form of (64) is

$$T + \frac{dU^+}{dy^+} = 1. \tag{66}$$

To summarize, we have the traditional approximation that asserts that $T$ is constant (but no approximation as yet for $U^+$) in the outer region near the channel's centerline, and an equation (64) or (66) relating $T$ and $U^+$ in the inner region.

The region next to the wall where the spatial variations (in $y$) have characteristic length $\frac{\nu}{u^*}$, is call the *wall layer* (as opposed to boundary layer). The choice (63) of scaling in this region says that we are treating the scaled Reynolds stress $T$ on the same footing as the wall stress or skin friction $(R^*)^{-1}\frac{dU^+}{d\eta}$. The former arises from the inertia terms in the Navier-Stokes equations and the latter from the viscosity terms.

In the wall layer, then, it is reasonable to suppose that $U^+$ and $T$ are (approximately) functions only of the inner variable $y^+$. This property is called the *law of the wall*.

## 5.5 Velocity in the core

The law of the wall is no longer valid for large $R^*$ as we move into the interior of the channel and on into the core region. The outer scaling will turn out to be valid there.

But remember, at this point we only have that $T$ is approximately constant there. $U^+$ may be unbounded as a function of $\epsilon$; the traditional way to incorporate the validity of the outer coordinate $\eta$ in an expression for $U^+$ is to postulate the *defect law*

$$U^+ = U^+(1) + h(\eta), \tag{67}$$

where $h$ is unknown, except that $h(1) = 0$.

The concept of thickness of the wall and the core scaling regions should be clarified. One interpretation of these regions would be where the solution's characteristic length is unity in the appropriate scaled coordinate. Then the thickness would be $O(1)$ as measured in that scaled variable, be it $y^+$ or $\eta$. In particular, the wall region is characterized by $\{y^+ \leq O(1)\}$, and that of the outer scaling region is characterized by $\eta_0 < \eta \leq 1$ for some arbitrary positive number $\eta_0$ independent of $\epsilon$. Note that the ratio of the thickness of the wall layer to that of the core is $O\left(\frac{1}{R^*}\right)$; this contrasts with the laminar hydrodynamic boundary layer (Prandtl theory), where the ratio is $O\left(Re^{-1/2}\right)$.

Another interpretation of the two regions would be where one or the other of the approximate representations of the solution, namely the law of the wall or the defect law, are valid. This interpretation yields, as it turns out, larger regions.

Let us assume that these two laws are correct in their respective domains; better justification for this will be given later in Sec. 6.8. Recall

$$\epsilon = (R^*)^{-1/2}; \tag{68}$$

since $U^+(1)$ will depend on $\epsilon$, we set $U^+(1) = G(\epsilon)$.

In the outer region, then, we may posit the representation

$$U^+ = G(\epsilon) + u_o(\eta). \tag{69}$$

In the inner region, on the other hand, we are using the law of the wall approximation

$$U^+ = p(y^+), \tag{70}$$

i.e. $U^+$ is a function only of $y^+$.

The unknown quantities here are $p$, $G$, $u_o$.

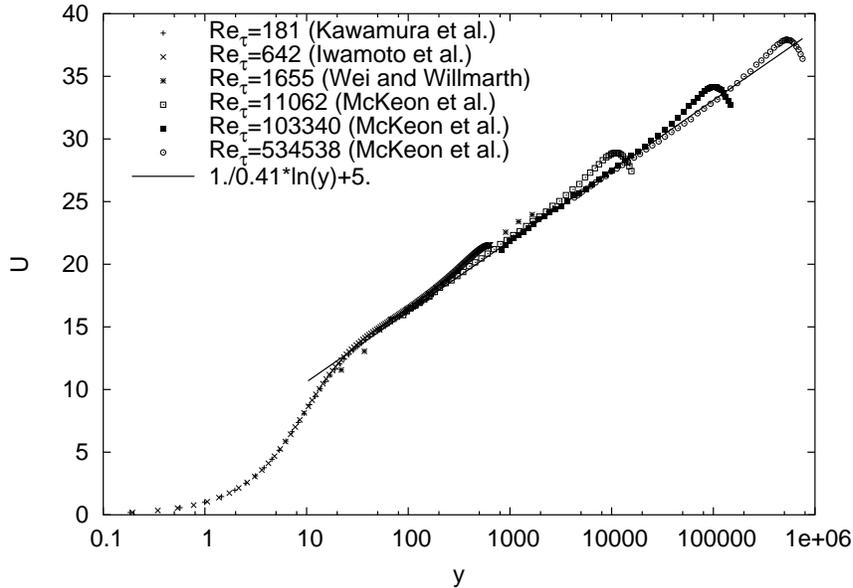

Figure 1: Inner normalized mean streamwise velocity in Couette flow, pressure-driven channel flow and pipe flow. Couette flow data are from DNS of Kawamura's group [13, 14]. Pressure-driven channel flow DNS data are from Iwamoto et al [15] and experimental data are from Wei and Willmarth [16]. Pipe flow data are from superpipe data of McKeon et al [17].

## 5.6 Application of the Izakson-Millikan reasoning

A great number of analytical and semianalytical studies of turbulent mean profiles have utilized the Izakson-Millikan observation as an essential ingredient. It has also been generalized and elaborated upon in many ways. For example the idea of composite expansions in more traditional settings of two-scale problems has been extended to the present scenario. We recommend the review paper [12].

Given the two approximations (69) and (70), they can be thought of as the inner and outer functions specified in the two sides of (40). The variables $p$, $u_o$, $G$, $y^+$, $\eta$ are analogous to $U$, $u_o$, $G$, $y$, $\eta$. Thus the left side of (40), $u_o(x) + G(\epsilon)$, is analogous to the right side of (69), and the right side of (40), $U(y)$, is analogous to the right side of (70).

If we now make the Izakson-Millikan hypothesis that there exists a common region in which the two expressions are almost equal, then the three functions in question must be either approximately constant or approximately logarithmic, as in (43) and (44).

In the present context, the conclusion is that in the common region, whatever it is, the following hold:

$$u_o(\eta) \approx A \ln \eta + B, \quad p(y^+) \approx A \ln y^+ + C, \quad G(\epsilon) \approx -A \ln \epsilon + C - B. \qquad (71)$$

Therefore the functions are approximately either constant ($A = 0$) or logarithmic.

This is a well-known conclusion, and indeed the mean velocity profile in wall-bounded turbulent flows is seen to exhibit logarithmic type behavior in certain regions which can be estimated on the basis of experimental data. Examples of the logarithmic property can be seen from the empirical data shown in Fig. 1. The coefficients $A$, $B$, $C$ can also be so estimated. All in all, the Isakson-Millikan observation, in all its simplicity, should be counted as one of the great success stories of theoretical turbulence.

Focussing on our stated objective to at least try to answer the question *why?*, we discuss the given derivation of the logarithmic property in some detail. In particular, we ask whether it can be supported by alternative trains of thought.

## 5.7 Observations on the foregoing procedure

The conclusion (71) gives a surprising amount of information about the inner and outer approximations, based on what appears to be a small amount of input.

The basis for the argument rests on very little physics or fluid dynamics; it is simply an assumption about inner and outer approximations agreeing somewhere. If one is willing to admit the existence of those inner and outer approximations, what remains is simply a mathematical issue, and could apply to any situation where there are two space scales with different but overlapping domains representing a strictly monotone function.

Let us rephrase what has been found in terms of the more realistic conclusion corresponding to Sec. 4.7. If the mean velocity profile is everywhere monotone and there is a region in the flow where that profile can be expressed approximately and simultaneously as a function of the inner variable alone and the outer variable alone (up to an additive function of $\epsilon$ alone), then these functions must be approximately logarithmic, the degree of the latter approximation being dependent on that of the former.

Let us take for granted the monotone part; that property of the mean velocity profile is well known and can be rationalized by the supposition that the viscous stress is everywhere positive and a decreasing function of distance from the wall, which is the site of the imposition of such stress by outside means. Given the monotonicity, what other information can we use to determine the profile, at least approximately? The needed additional information should be theoretical in nature, because our aim is to explain the reasons for observed behavior. We know the Izakson-Millikan implication, which, stated succinctly, says, "overlap implies logarithmic". It is a simple piece of reasoning which is classical and, again, well-known. Being so simple and direct, it places the hypothesis that a particular region is an overlap region very close to the conclusion, namely close to assuming logarithmic behavior. Either property can be substituted for the other. If this is true, the argument is close to being circular. If available, independent arguments to determine properties of the mean velocity and Reynolds stress profiles, not relying on the overlap hypothesis, would be highly desirable. That will be a principal aim in the following sections.

One direction of enquiry which is suggested is to consider all possible increasing profiles which are functions of the inner variable near the wall and of the outer variable near the centerline; and try to use some reasonable selection criterion to at least make a good argument for what the actual profile should be. The set of profiles which have the overlap property, hence the logarithmic property, is only a small part of the set of all conceivable profiles, because the latter includes an ample collection of arbitrary non-logarithmic ones. If the overlapping logarithmic ones are to be selected, as empirical evidence suggests in some regions, then theory should be able to supply a reason for that choice.

We should mention in passing a small cloud hanging over the search process: it seems on the basis of examples explored in Sec. 4 that overlap regions, if they exist, are usually characterized by the functions being constants in those regions. Given that $U$ is nowhere constant, one wonders whether the assumption that an overlap region exists is itself reasonable.

Another possible direction of enquiry is to ask whether a unified approach is possible, which gives more justification to the inner and outer scalings themselves, and at the same time is capable of handling all other parts of the profile as well.

In the next subsection we draw attention to an intuitive and vague train of thought which may be relevant to the question of why the mean profiles might have certain features. In Section 6, a relatively new approach to this and other questions will be explored. It involves a systematic search for scaling patches and does not require separate assumptions about the validity of inner and outer scales.

Apart from the above, another consideration to be kept in mind is that the overlap hypothesis provides no theoretical basis for determining the location of any expected overlap region, nor the accuracy of (69) in that region (neither does the scaling patch approach, for that matter). It is expected that the size of the region is related to the accuracy of the basic hypothesis. Increasing the region where the two expressions are approximately valid decreases the accuracy of the approximation in that region.

## 5.8 The uniformizing effect of turbulence and some possible implications.

The turbulent nature of the fluid motion, which is not explicit in the averaged equation under study, has a mixing and hence uniformizing effect. This, it may be argued, tends to smooth out or eliminate abrupt spatial changes in the average properties of the flow as one moves from one location to another. The flow properties at one place will be similar to what they are at nearby places.

Although sources of stress which are imposed from the outside, such as wall friction and enjoined pressure gradients, are often also sources of turbulence, they may work in opposition to the uniformizing action. For example the generation of stress due to the wall in channel flow acts at a different location from an imposed pressure gradient, and this difference results in a nonuniform distribution of Reynolds and viscous stress.

Still, the claim of uniformization seems without much doubt to be valid for some properties if the vague terms "properties" and "similar" are given some appropriate definition.

If we assume that this principle holds for the local scaling properties of the variables, including for the local characteristic length as it depends on location, then there should be a whole continuum of characteristic lengths, each appropriate to a specific locale, i.e. specific distance from the wall. Thus as we pass from the wall to the centerline of the channel, the uniformizing principle suggests that the characteristic length encountered during the passage should change continuously from the one for the inner scaling to the one for the outer scaling.

In fact, this is one of the main conclusions to be brought out by a different reasoning in the next section, which introduces and develops a very different approach to the problem of determining the nature of the profiles. It uses scaling and order of magnitude arguments in a major way, but does not assume a priori the validity of the inner or outer approximations. Rather, it presents an alternative criterion for "scaling patches" and shows that the inner and outer regions, among many others, fit that criterion. In this sense, the identification of patches is derived from the assumed criterion rather than presupposed.

## 6 The search for scaling patches

The more recent approach to understanding of the scaling structure of the mean velocity and Reynolds stress profiles presented here was introduced in [7, 8, 9, 10, 11]. It forms an alternative to the approaches considered above, not in any sense of replacing them, but rather in the spirit of adding to them new information and insights. Finally, it is applicable to a wide variety of wall-induced turbulence scenarios. For the time being, we continue to operate within the context of turbulent Couette flow through a channel.

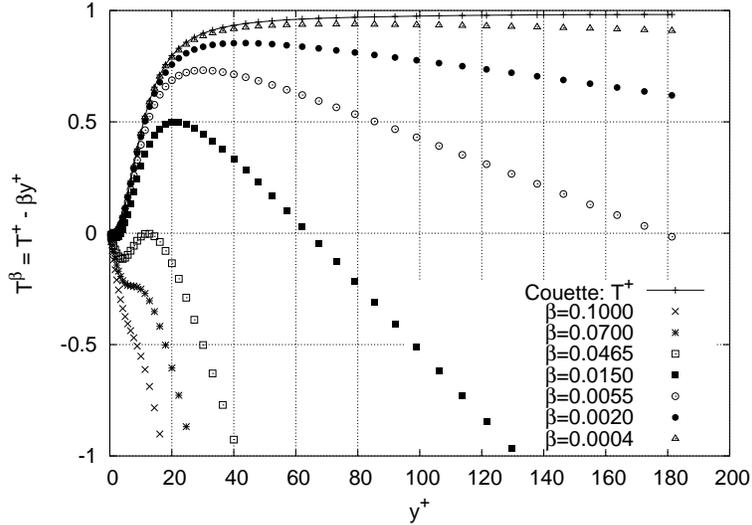

Figure 2: Adjusted Reynolds stress profile for various values of $\beta$. The case $\beta = \epsilon^4$ corresponds within $O(\epsilon^2)$ to the genuine Reynolds stress for Couette flow and $\beta = \epsilon^2$ is an approximation to that for pressure driven channel flow. The DNS data are from [18], $\delta^+ = Re_\tau = 181.3$ and $\epsilon = .074$.

### 6.1 Adjusted Reynolds stresses, balance exchange phenomena, and the identification of patches

Let $\beta$ be a small positive number. Restrictions on it will be given later. In terms of the originally defined dimensionless Reynolds stress $T$, let

$$T^\beta(y^+) = T(y^+) - \beta y^+. \tag{72}$$

(Note that $\beta$ is a superscript not an exponent.) The function $T^\beta$ is simply a mathematical construct that will be called an adjusted Reynolds stress. It satisfies

$$\frac{dT^\beta}{dy^+} = \frac{dT}{dy^+} - \beta, \tag{73}$$

and from (64)

$$\frac{d^2 U^+}{dy^{+2}} + \frac{dT^\beta}{dy^+} + \beta = 0. \tag{74}$$

The $T^\beta$ are plotted for various values of $\beta$ in Fig. 2. The genuine Reynolds stress, which corresponds to the case $\beta = 0$, vanishes with its derivative at the wall and rises to attain a maximum at the centerline. As $\beta$ increases, the position of the maximum moves toward the wall, eventually disappearing. The main interest is in those adjusted stress functions that exhibit local maxima—which will be the case when $\beta$ is not too large—because, as it turns out, scaling patches $L_\beta$ with characteristic lengths with orders of magnitude $O(\beta^{-1/2})$ (as $\beta \to 0$) exist at those peak locations. The reasoning below will justify this assertion. Part of the argument involves obtaining an exact differential equation in rescaled variables having no explicit dependence on $\epsilon$ or $\beta$. Another part entails the recognition that (74) expresses an approximate balance between its first two terms (since $\beta$ is small), and that this balance is necessarily broken at some point and changed to another kind of balance, because $y^+$ eventually attains a value such that the three terms in (74) have the same order of magnitude.

Let us pursue this idea of balance exchange. As was brought out in Sec. 5.2 following (59), the function $T(y^+)$ increases from being 0 at the wall to attain its maximal value at the centerline $\eta = 1$, i.e. $y^+ = \epsilon^{-2} = \delta^+$. Assuming that $\beta$ is small and positive, therefore, we see that the adjusted stress $T^\beta$ has negative derivative at $y^+ = \delta^+$, so must attain its maximal value $T_m(\beta)$ at a point $y^+_\beta < \delta^+$. Moreover, the location of this maximum decreases as $\beta$ increases, because when $\beta$ increases, the zero derivative at any maximum becomes negative.

Within the inner scaling region where the law of the wall holds, for example when $y^+ \leq O(1)$, the two derivatives in (74) will generally have magnitudes $O(1)$ except very near the wall in the viscous sublayer, where they are both very small. Since $\beta \ll 1$, those two derivatives will balance, except for an error represented by the last term in (74). Both derivatives will therefore generally be $O(1)$ quantities, except as noted above. This occurs within the inner scaling region. However as $y^+$ increases to a neighborhood of $y^+_\beta$, this necessarily changes, because the value of $\frac{dT^\beta}{dy^+}$ decreases to zero at $y^+ = y^+_\beta$. For points near enough to that value, the second term in (74) must take on values $\leq O(\beta)$, and therefore by (74) again, the first term does as well. It is therefore natural to propose that there may be a scaling patch occupying that neighborhood with respect to which all three terms of (74) have the same formal order of magnitude. To be more precise, the last term balances the sum of the first two terms, each of which is $\leq O(\beta)$.

It turns out that it is possible to construct a candidate for such a patch. It will be centered at the location $y^+ = y^+_\beta$, where $\frac{dT^\beta}{dy^+} = 0$. At that point, the derivatives appearing in (74) annihilate the linear terms in the Taylor series of the functions $U^+$ and $T$ about $y^+ = y^+_\beta$ (in fact the linear term in $T$ is identically zero). As a result, those linear parts do not play a role in the rescaling process, and one may work only with the remainders after those parts have been separated off. With this in mind, we write, for some coefficients $\alpha(\beta)$, $\gamma(\beta)$, $\lambda(\beta)$ to be determined,

$$y^+ = y^+_\beta + \alpha \hat{y}, \quad T(y^+) = T_m(\beta) + \gamma \hat{T}(\hat{y}), \quad U^+(y^+) = U^+(y^+_\beta) + m(y^+ - y^+_\beta) + \lambda \hat{U}(\hat{y}), \qquad (75)$$

where $m$ is the slope $m(\beta) = \frac{dU^+}{dy^+}(y^+_\beta)$ of the mean velocity profile. The slope $m$ is unknown at this point, but will be found later in Sec. 6.4.3. The linear parts, which have been segregated in (75), are $T_m(\beta)$ and $U^+(y^+_\beta) + m(\beta)(y^+ - y^+)$, respectively. They are separated off because the derivatives appearing in (74) annihilate them and they take no part in the present calculation. That is also why $m$ is not determinable until later in Sec. 6.4.3 (see (93)). The new rescaled variables are $\hat{y}$, $\hat{T}$ and $\hat{U}$. Then (74) becomes

$$\frac{\lambda}{\alpha^2} \frac{d^2 \hat{U}}{d\hat{y}^2} + \frac{\gamma}{\alpha} \frac{d\hat{T}}{d\hat{y}} + \beta = 0. \qquad (76)$$

In order for the three terms to be formally equal in order of magnitude, one can specify

$$\alpha = \left(\frac{\lambda}{\beta}\right)^{1/2}, \quad \gamma = (\lambda \beta)^{1/2}, \qquad (77)$$

so that

$$y^+ = y^+_\beta + \left(\frac{\lambda}{\beta}\right)^{1/2} \hat{y}, \quad T(y^+) = T_m(\beta) + (\lambda \beta)^{1/2} \hat{T}(\hat{y}), \quad U^+ = U^+(y^+_\beta) + my^+ + \lambda \hat{U}(\hat{y}), \qquad (78)$$

and (74) is transformed into the parameterless equation

$$\frac{d^2\hat{U}}{d\hat{y}^2} + \frac{d\hat{T}}{d\hat{y}} + 1 = 0. \tag{79}$$

The criterion of equal formal orders of magnitude, therefore, does not by itself determine uniquely the three scaling factors $\alpha$, $\gamma$, $\lambda$; (77) leaves the factor $\lambda$ undetermined. This suggests that there may be a one parameter family of potential scaling patches at the location $y_\beta^+$, the parameter being $\lambda$. We are confronted with an extra degree of indeterminacy, because the present line of reasoning does not offer a way to determine which of the potential patches represent actual ones. However there is considerable evidence, to be summarized in Sec. 6.7, that the correct scaling at this location $y_\beta^+$ is given by (77) with $\lambda = 1$. The analysis to follow in this and the next two subsections holds for other choices of $\lambda$ as well.

We assume that $\lambda(\beta)$, like $\alpha$ and $\gamma$, is a power of $\beta$, and define, in place of $\lambda$, the parameter $\sigma$ by $\lambda = \beta^{-\sigma}$. (No constant coefficient is needed with this power law because we are dealing only with orders of magnitude.) It will be shown that the case $\sigma = 0$ leads to a logarithmic-like profile for the mean velocity $U^+$, and when $\sigma$ is positive, we get behavior like a power law with exponent depending on $\sigma$.

For reference, the prototypical case is $\sigma = 0$, when

$$y^+ = y_\beta^+ + \beta^{-1/2}\hat{y}, \quad T(y^+) = T_m(\beta) + \beta^{1/2}\hat{T}(\hat{y}), \quad U^+ = U^+(y_\beta^+) + m(\beta)(y^+ - y_\beta^+) + \hat{U}(\hat{y}). \tag{80}$$

No information is available at this point about the slope $m$, which bears on the profile of $U^+$. However, information about it will be found later (93).

We argue that the scaling (78), for some choice of $\sigma$, is the natural one in a neighborhood of $y_\beta^+$, and that this neighborhood is a scaling patch. On the basis of the explanation in Sec. 4.1 together with (75) and (77), it will follow that the characteristic length in that patch will be

$$\ell(\beta) = \left(\frac{\lambda}{\beta}\right)^{1/2} = \beta^{-(\sigma+1)/2}. \tag{81}$$

In fact not only does the scaling produce a parameter-free exact form (79) of the momentum balance equation, but at locations in the proposed patch it can be verified that the individual derivatives in (79) have the right order of magnitude, namely $\leq 1$ with at least some of them $= O(1)$). For example, at the peak location, the three terms on the left of (79) are $-1$, $0$, $1$ respectively. Leading up to that peak, the middle term is positive but still $\leq O(1)$, which makes the first term also $O(1)$, according to (79) again.

In a scaling patch, as defined originally, all derivatives using the scaled variables are $\leq O(1)$, and in the case of at least one of those variables, the magnitudes of its derivatives are not all strictly $< O(1)$. If that were not the case, the scaling factor $\alpha$ could be decreased without forcing some of the new rescaled derivatives to be unbounded as $\beta \to 0$. In the present case, we have shown that these order of magnitude relations hold for the particular derivatives appearing in (79). That fact makes the scaled neighborhood of $y_\beta^+$ with width $O(1)$) in $\hat{y}$ a candidate for a scaling patch, provided the correct choice of $\sigma$ is taken. This will be our accepted criterion for the existence of a patch.

Thus if the correct $\sigma$ is taken, then given any suitable small number $\beta$, one concludes that there is a corresponding scaling patch, which we shall call $L_\beta$, with characteristic length $\ell(\beta) = \beta^{-(1+\sigma)/2}$ located at the point $y_\beta^+$ where $T^\beta$ achieves its maximum.

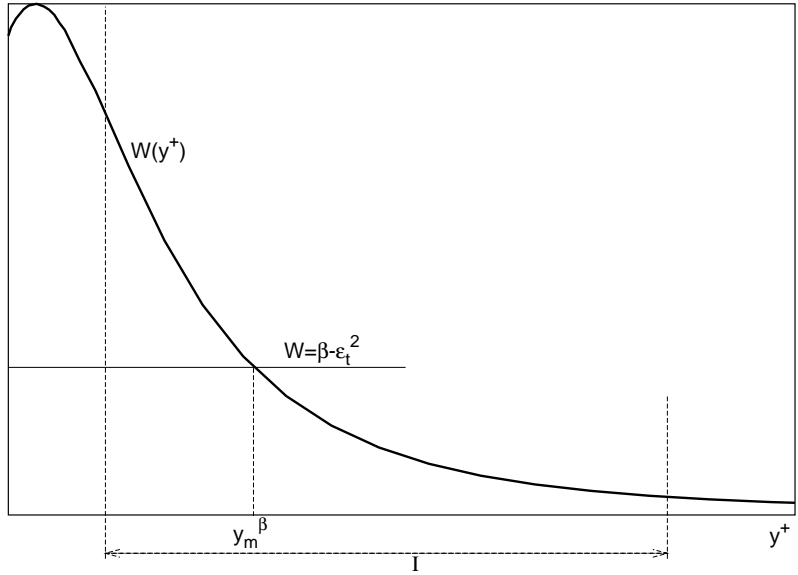

Figure 3: Schematic diagram showing the role of the function $W(y^+)$, which is called $P(y^+)$ in the text, in determining the relation between $\beta$ and the location $y_m^\beta$ of the corresponding scaling patch. The interval $I$ shown here is one of many choices of interval on which $W(=P)$ is decreasing. See (82).

One may question whether this characteristic length $\ell$, for the correct choice of $\sigma$, is comparable to the mixing length of Prandtl in Sec. 3 or the scaling parameter introduced by von Karman (15). They arise from apparently different considerations; the lengths used by von Karman and Prandtl are characteristic lengths vaguely associated with the fluctuating velocity, simply postulated to exist, whereas the present one is related to the function $T$. And yet $T$ is in essence something which is defined in terms of those fluctuations, and so one may surmise that the two concepts are related somehow. Because of the vagueness of mixing length concepts, that may be all that can be said.

## 6.2 The locations of the scaling patches

In the following the analysis will be done only for the case $\sigma = 0$. Analogous calculations may be done in the other cases as well; we will simply provide some key equations in the general case, without derivations.

At this point one should ask how to determine the range of parameter values $\beta$ for which the foregoing construction of scaling patches is possible. This is best answered at first in terms of the known qualitative properties of the function $P(y^+) \equiv \frac{dT}{dy^+}(y^+)$; later a more complete answer will be given. The function $P$ vanishes at the wall ($y^+ = 0$) and, by symmetry of the function $T$, also at the centerline $y^+ = \delta^+ = \epsilon^{-2}$. As $y^+$ passes from the wall to the centerline, $T$ rises to its maximum at the latter location; during the transition, $T$ and $P$ are both positive. Since $P = 0$ at those two locations, it must attain a positive maximum at some intermediate point; call it $y_p^+$. Being the gradient of $T$, $P$ is expected take on its greatest values in the inner region, where the length scale is shortest and gradients are largest. Therefore the location $y_p^+$ of the positive maximum of $P$ will be expected to lie in the inner region, so that $y_p^+ = O(1)$.

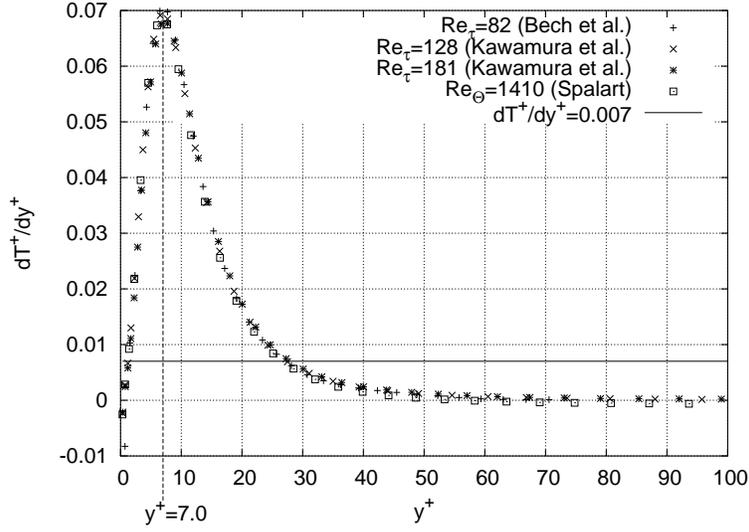

Figure 4: Inner normalized Reynolds stress gradient for a variety of flows. The turbulent Couette flow data are from [19] and [18]. Also included are the turbulent channel DNS data from [20] and turbulent boundary layer DNS from [21].

For $y^+ > y_p^+$, $P(y^+)$ will decrease from its maximum. Let us assume it is a decreasing function on the entire interval $(y_p^+, \delta^+)$. Given any point $y^*$ in that interval, let $\beta(y^*)$ be the corresponding value of $P$, i.e.

$$P(y^*) = \beta. \tag{82}$$

According to (73), which can be written $\frac{dT^\beta}{dy^+} = P(y^+) - \beta$, the function $\frac{dT^\beta}{dy^+} = 0$ at the point $y^*$ under consideration, and since $P$ is a decreasing function, $\frac{dT^\beta}{dy^+}$ changes from being positive to being negative as $y^+$ increases past the point $y^*$. Therefore $T^\beta$ has a maximum, which we may call $T_m^\beta$, at that point. The point $y^*$ in question may therefore also be labeled $y_\beta^+$. This is illustrated in Fig. 3, with plots of typical functions $P$ in Fig. 4.

What has been shown is that any value $y^* \in (y_p^+, \delta^+)$ will serve as $y_\beta^+$ if $\beta$ is chosen to be $P(y^*)$. The set of all such values of $\beta$ is the set for which the above construction of a scaling patch will work, namely the set of all values of $P(y^+)$ for $y^+ \in (y_p^+, \delta^+)$. That range in $y^+$ provides, then, the locations where we have succeeded in finding a scaling patch; moreover we have found the characteristic lengths of all these patches; they are given according to (80) in the case $\sigma = 0$ by $\beta^{-1/2}$. These characteristic lengths increase with increasing distance from the wall, since $P$ decreases with distance. It will be argued in Sec. 6.4, in fact, that asymptotically as $\beta \to 0$, they are proportional to distance from the wall.

### 6.3 More on the locations of the patches

It will be desirable to correlate the locations $y_\beta^+$ of the scaling patches with the characteristic lengths given by (81). The above discussion accomplishes this in terms of the function $P$. We now ask whether information can be obtained even if $P$ is not known.

In order to proceed we now further exploit the facts that $T^\beta$ has a maximum at $y^+ = y_\beta^+$, and that (80) expresses the normal (natural) scaling near that point. In terms of the rescaled variables given in (80), we have that $\hat{T}$ has a maximum of 0 at $\hat{y} = 0$. We are going to emphasize the variation of $\beta$ and the dependence of the rescaling shown in (78) on $\beta$. Then the fact that $T^\beta(y^+)$ has a maximum at $y_\beta^+$ implies that for all $\beta$ in the allowed interval of values,

$$\frac{dT^\beta(y^+)}{dy^+}(y_\beta^+) = \frac{dT(y^+)}{dy^+}(y_\beta^+) - \beta = 0. \tag{83}$$

This may be differentiated with respect to $\beta$ to yield

$$\frac{d^2 T(y^+)}{(dy^+)^2}(y_\beta^+)\frac{dy_\beta^+}{d\beta} - 1 = 0. \tag{84}$$

Writing the second derivative on the left in terms of the scaled variables, we find

$$\beta^{3/2} \left.\frac{d^2 \hat{T}(\hat{y})}{(d\hat{y})^2}\right|_{(\hat{y}=0)} \frac{dy_\beta^+}{d\beta} - 1 = 0. \tag{85}$$

In the patch $L_\beta$, the scaled variables satisfy (79), which is parameter-independent; moreover $\hat{T}$ satisfies the two equations $\hat{T}(0) = \frac{d\hat{T}}{d\hat{y}}(0) = 0$, which are also parameter-independent relations. One can therefore argue that the second derivative on the left of (85), which we shall designate by

$$-A \equiv \left.\frac{d^2 \hat{T}(\hat{y})}{(d\hat{y})^2}\right|_{(\hat{y}=0)}, \tag{86}$$

should not depend in any major way on the parameter $\beta$. This is reminiscent of similarity hypotheses, since an assumption that $A$ is constant is an assumption that the quantity $A$ is invariant under certain transformations associated with changing $\beta$. Certainly the order of magnitude of $A$ (with respect to $\beta$)) is $O(1)$, hence $\beta$-independent, and we shall argue below in Sec. 6.6 that in some regions any $\beta$-independence should vanish in the limit as $\epsilon \to 0$. From (85), then,

$$\frac{dy_\beta^+}{d\beta} = -A^{-1}\beta^{-3/2}. \tag{87}$$

Recall that $y_\beta^+$ is the location, in the original inner variable, of the scaling patch with characteristic length $\beta^{-1/2}$. The equation (87), therefore, provides some insight into the dependence of that characteristic length on location, without using much knowledge about the function $P$.

### 6.4 The case $A = $ constant

If $A$, as it depends on $\beta$, were known, the locations $y_\beta^+$ could be found by solving differential equations. It is not known, of course; but its order of magnitude is known to be unity. That knowledge provides order of magnitude information about the profiles and about the locations $y_\beta^+$. It is most instructive at this point to look at the calculations in the easiest case $A = $ constant. After that, we can see how the results so obtained are still valid in an order of magnitude sense.

#### 6.4.1 Characteristic length as it depends on location

As mentioned, the scaling given in (78) shows, among other things, that the characteristic length $\ell$ of a patch is given by (81). We shall sometimes write relations below in terms of $\ell$ instead of $\beta$.

The symbol $C$, with or without subscript, will denote a variety of different constants, sometimes integration constants, which depend on $\sigma$ but not on $\beta$ or $A$.

If $A$ is constant in some given interval, then in that interval (87) may be integrated to obtain

$$y_\beta^+ = -A^{-1} \int \beta^{-3/2} d\beta = A^{-1} C_0 \beta^{-1/2} + C_1 = A^{-1} C_0 \ell + C, \tag{88}$$

where the final $C$ is an integration constant. We now drop the subscript $\beta$ from $y_\beta^+$ and obtain the relation

$$\ell = CA(y^+ - C). \tag{89}$$

This same relation in fact holds as well for other choices of $\sigma$. The left side is the characteristic length of the patch located at $y^+$; except for a shift in the independent variable $y^+$, that length is proportional to distance from the wall.

### 6.4.2 The Reynolds stress

In (82) the point $y^*$ may be identified with the general point $y^+$ in (89), the location of the patch $L_\beta$. Substituting the expression for $\beta$ given in (89) into the right side of (82) and recalling the definition of $P$, we obtain

$$\frac{dT}{dy^+} = \beta = C_1 A^{-2} (y^+ - C_2)^{-2}. \tag{90}$$

Integrating this provides $T$ as a function of $y^+$:

$$T(y^+) = -C_3 A^{-2} (y^+ - C)^{-1} + C_4, \tag{91}$$

with another integration constant $C_4$. The relation (91) holds only under the assumption that $A$ is constant, and only for values of $y^+$ in the interval where $T$ is a decreasing function, because that is where the scaling patches were found. This range extends up to the channel midline at $y^+ = \delta^+$, at which point the left side of (91) vanishes. This gives a relation among $\delta^+$ and the integration constants. Since the right side of (90) cannot vanish and the left side does, the assumption that $A$ is constant is incorrect at least near the centerline.

### 6.4.3 The mean velocity profile

From (66) and (91) we find

$$\frac{dU^+}{dy^+} = 1 + CA^{-2}(y^+ - C)^{-1} - C_5. \tag{92}$$

We require this derivative to be small for large $y^+$; this simply means that at the centerline $U^+$ must become flat as $\epsilon \to 0$. Therefore it appears that $C_5 \approx 1$ and

$$\frac{dU^+}{dy^+} \approx CA^{-2}(y^+ - C)^{-1}. \tag{93}$$

Combining (93) with (89) yields an asymptotic relation between the characteristic length $\ell$ and the slope of the mean velocity profile:

$$\frac{dU^+}{dy^+}(y^+) \approx CA^{-2} \ell^{-1}. \tag{94}$$

For other values of $\sigma$ the right side should be replaced by $CA^{-2/(1+\sigma)} \ell^{-(1-\sigma)/(1+\sigma)}$. In any case, we may express this in terms of $\beta$ and identify it with the slope $m$ in (78). Note that in the case $\sigma = 0$, $m \approx \ell^{-1}$, which says that the characteristic length of the linear part of $U^+$ in (78) is the same, $\ell$, as that of the nonlinear part. More generally, in any case if the latter is true, then $m$ should also be identified with the ratio of characteristic increments $\Delta U^+/\Delta y^+$, which from (78) is $\lambda/\alpha = \beta^{(1-\sigma)/2} = \ell^{-(1-\sigma)/(1+\sigma)}$. This is in agreement with (94). The conclusion is that for any choice of $\sigma$, the characteristic slope of the $U^+$ profile calculated by integrating (87) on the basis of the nonlinear increments is the same as that which one would surmise by assuming that the characteristic increments for the linear parts of (78) are the same as for the nonlinear parts. This fact lends credence to that assumption.

Integrating (93) again gives

$$U^+ \approx \begin{cases} CA^{-2/(1+\sigma)}(y^+ - C)^{2\sigma/(\sigma+1)} + C_6, & \sigma > 0, \\ CA^{-2}\ln(y^+ - C) + C_6, & \sigma = 0. \end{cases} \quad (95)$$

This expression (95) in the case $\sigma = 0$ is similar to that (71) for the mean velocity in a hypothesized overlap region given by the Izakson–Millikan argument. In the case $0 < \sigma \ll 1$ it gives a power law with small exponent; that also has been suggested in the past, but experimental or DNS data has left the resolution of the question unclear.

In both the Izakson–Millikan argument and the present argument, questionable assumptions lead to the derivation of (95), or at least part of it. In the I-M case, those assumptions have been reviewed in Section 4.6. In the present case it is mainly, but not solely, the assumption that $A$ is constant. The crucial assumption in either case must realistically be only approximate, with unknown error. In the present scenario, however, there is a good reason to believe that in some regions, the error in the assumption about $A$ approaches 0 as $\epsilon \to 0$, i.e. in the limit of large Reynolds numbers. That reasoning will be given below in Sec. 6.6.

## 6.5 Relaxing the assumption that $A$ is constant

Although the similarity assumption $A \sim 1$ is true in order of magnitude, it is unlikely to be strictly true except in interior regions at high Reynolds numbers, as shown below in Sec. 6.6.

If $A = A(\beta)$ depends on $\beta$, then (87) is still valid. Moreover if $A$ remains $O(1)$, i.e. bounded above and below by positive constants independent of $\beta$ or $\epsilon$, then (88) and (91) still hold in a weakened sense. They are replaced by pairs of inequalities for some constants $K_i$ independent of all parameters. In the logarithmic case $\sigma = 0$, we have

$$K_1 \ell \leq y_\beta^+ \leq K_2 \ell, \quad (96)$$

$$K_3(y^+ - C)^{-2} \leq \frac{dT}{dy^+} = \beta \leq K_4(y^+ - C)^{-2}. \quad (97)$$

Thus in an asymptotic sense as $y^+ \to \infty$, the characteristic length is still proportional to distance from the wall.

Similar transformations to pairs of inequalities are valid for (90)–(95). For example if $\sigma = 0$ the latter becomes

$$K_5 \ln y^+ \leq U^+ \leq K_6 \ln y^+. \quad (98)$$

In domains where $A$ is nearer to being constant, such inequalities are valid with constants $K_i$ which are closer to one another. In fact the error in assuming $K_5 = K_6$, for example, can be estimated in terms of any assumed error bound in assuming that $A =$ constant in the domain being considered. This is parallel to the error consideration in the Izakson-Millikan argument, which was detailed in Sec. 4.7. In both cases the conclusion of logarithmic growth is probably never exact; it should be considered an approximate statement, with the accuracy of the statement dependent on the accuracy of the underlying assumption. In the I-M case, the underlying assumption is that the outer and inner approximations are both exact in some domain; in the present case, it is that the quantity $A$ is constant in some domain. In both cases there is no theoretical way to gauge how accurate the approximations are (see, however, the following subsection).

## 6.6 Approximate constancy of $A$ in interior zones

As was brought out before, the order of magnitude of the quantity $A$ remains $O(1)$ in $\beta$ and $\epsilon$. In locations far away from the endpoints of the range of the continuum of scaling patches, it can be argued that $A$ should be almost constant. The reason is that the data we have for $A$, namely the differential equation it satisfies and the known exact values of the terms in that equation, are parameter-independent. Therefore any variation in $A$ due to changes in $\beta$ will be caused not from those sources. This invariance as $\beta$ changes suggests, by a similarity consideration, that $A$ is constant if it is not subject to other influences. Those would be influences from neighboring patches, hence ultimately from locations, on either side of the continuum, where the boundary would introduce "external" influences. Only at those places would the similarity suffer external disruption. And the effects of that disruption would be most likely to happen near those disrupting sites, either toward the outer or the inner zones. That leaves interior regions far away from those zones as candidates for places where $A$ is nearly constant. It was shown above in Sec. 6.4.3 that those are the regions where the mean velocity profile is logarithmic-like.

The extent (in inner units) of these zones of near similarity will grow as $\epsilon$ becomes smaller, because there will be a larger range of patches far away from the extremal patches.

## 6.7 Evidence for logarithmic growth, i.e. $\sigma = 0$

The possibility that the mean velocity profile grows in parts of the flow according to a power law, rather than a logarithmic law, has been discussed by other authors. In either case, the actual expression for the mean velocity will depend somewhat on the Reynolds number and is very unlikely to be exactly logarithmic or a power function. These may represent approximations, but that is all. What we are concerned with are trends, brought about by relations such as (98) and its analog for power functions. They are generated by the scaling parameter $\lambda = \beta^{-\sigma}$ in (78). Here we summarize the evidence in favor of choosing $\sigma = 0$.

- The Izakson-Millikan argument (Sec. 5.6).

- It was shown from empirical data in [7] that the increment in $U^+$ across the mesolayer is $O(1)$, in fact near 1 independently of the (large) Reynolds number. The mesolayer is one example of a scaling patch for turbulent Poiseuille flow. The analysis of that flow in Sec. 7 follows the present analysis (which has been for Couette flow). The role of the parameter $\sigma$ in determining the increment in $U^+$ across any scaling patch, including the mesolayer, is seen from (75). The nonlinear part of the increment is simply $O(\lambda) = O(\beta^{-\sigma})$, and it was brought out following (94) that in all cases the linear part of the increment, governed by $m$, is the same. Therefore the only case in which this increment is $O(1)$, as apparently required in the mesolayer, is the case $\sigma = 0$.

## 6.8 The inner scaling patch at the wall and the outer scaling patch at the midline

The construction of the patches given in section 6 has, as a primary ingredient, the fact that as the peak in the adjusted Reynolds stress is approached, a region must appear in which all three terms in the mean momentum balance equation, which in this case is (74), will have the same order of magnitude. This is simply because the gradient $\frac{dT^\beta}{dy^+}$ approaches 0. (There will, of course, be a smaller region encompassing the peak in which the last term on the left of (74) has smaller order of magnitude than the others, because it vanishes at the peak.)

Curiously, a somewhat similar phenomenon happens when $y^+ \to 0$, since the gradient of the actual Reynolds stress, rather the adjusted one, is zero at the wall ($y^+ = 0$) and positive for small values of $y^+ > 0$. We are speaking of the first term $\frac{dT}{dy^+}$ in (64). With the inner scaling, both terms in (64) have equal orders of magnitude, both actual and formal. This in itself provides evidence that this wall region, together with the inner scaling used in (64), defines a scaling patch. But there is further evidence from the boundary condition (65) for $U^+$. That condition of course was engineered by our very choice of inner scaling. But whatever its origin, it furnishes decisive evidence that a scaling patch exists there. As before, the width of this patch is $O(1)$, measured in $y^+$. As was brought out before, it also encompasses the crucial point $y_p^+$ discussed in Sec. 6.2.

To summarize, at the wall, $U^+$, $T$, and $\frac{dT}{dy^+}$ are all 0, but $\frac{dU^+}{dy^+} = 1$ (that is how the inner scaling was selected). Thus all derivatives of interest in the scaled variables at that point either vanish, or (in one case) are unity. This circumstance is an adequate criterion for the validity of that scaling.

On the other end of the continuum, where $\ell$ is large, we know that $\ell$ reaches a maximum of $\epsilon^{-2}$, because that is the half-width of the channel and no larger scaling patch could fit into the latter. According to (81), it corresponds to $\beta = \epsilon^{4/(1+\sigma)}$. That forms a lower bound on the possible values of $\beta$. In the case $\sigma = 0$ for example, when $\beta = \epsilon^4$ the existence of a scaling patch can be ascertained by the previous argument in Sec. 6.1, which still holds true ($y_\beta^+$ for that value of $\beta$ must lie a distance $\leq O(1)$ from the centerline).

It should be noted that this "outer" patch encompassing the centerline is not the same as the traditional outer length scaling spoken of in Sec. 5.3, although the two ideas are compatible. The present concept of outer patch is that of an interval in the core together with a rescaling of all the variables, not just $y^+$, which will produce a parameter-free version of the mean momentum balance, namely (79), at that location.

Thus our construction of scaling patches is valid up to and including the centerline, and down to the inner region. In all, the scaling patches cover the entire channel.

## 6.9 Discussion

We have found that at each point in the Couette flow, the Reynolds stress and mean velocity profiles have a natural scaling; in other words, a scaling patch is located at that point. In order of magnitude, its characteristic length, which is essentially the width of the patch, increases continuously from 1 (in inner units) in the inner region where the law of the wall holds, to $\epsilon^{-2}$ at the centerline of the channel, where the outer scaling holds. Each patch is associated with a peak in one of the adjusted Reynolds stresses, and with a balance exchange event that occurs there involving that same adjusted stress.

The patches can be parameterized by their characteristic lengths, which increase monotonically with distance from the wall. Asymptotically as $y^+ \to \infty$ (since $y^+$ is limited by $\epsilon^{-2}$, necessarily $\epsilon \to 0$), the characteristic length is proportional to that distance, and in any case up to order of magnitude, is given by a solution of an ordinary differential equation.

Again up to order of magnitude, the mean velocity and Reynolds stress profiles are determined. In certain regions in the limit as the Reynolds number approaches $\infty$, these order of magnitude results are replaced by explicit functions. This is really a statement about the validity of an approximation for large Reynolds number.

The argument can be framed as a similarity and invariance issue. There is a family of rescalings depending on a parameter $\beta$, applied at $\beta$-dependent locations, which leave the governing rescaled momentum balance equation and associated numerical values of the derivatives appearing in that equation invariant as $\beta$ is varied. The statement then is that another derivative, denoted by $A$, is, in order of magnitude, invariant as well, and in some regions in fact approximately numerically invariant.

These results are consistent with logarithmic growth predicted by the Izakson–Millikan argument, which by hypothesis is to hold in some overlap zone between the inner and outer regions. The scaling patch and order of magnitude results presented here, on the other hand, are true throughout the channel; they generally differ from the exact logarithmic law except where and if the quantity $A$ is constant. Finally, this constancy condition is unlikely to hold exactly anywhere, except in the limit as $\epsilon \to 0$.

# 7 Turbulent Poiseuille flow in a channel

In this idealized picture, the two walls at $y = 0$, $y = 2\delta$ are stationary, so that their motion no longer provides impetus for the flow; however such an impetus is provided by a given pressure gradient streamwise along the channel. The treatment here follows that in [8, 9].

## 7.1 Differences from Couette flow

Mathematically, the differences between Couette and Poiseuille flows lie in the facts that $P_x$ is no longer 0 in (51), and the boundary conditions at the two walls are changed. Namely,

$$U = \langle u'v' \rangle = \frac{d}{dy}\langle u'v' \rangle = \frac{d^2}{dy^2}\langle u'v' \rangle = 0 \tag{99}$$

at $y = 0$ and $2\delta$; and at the centerline $y = \delta$ by symmetry,

$$\frac{dU}{dy} = \langle u'v' \rangle = 0. \tag{100}$$

In fact $U(y)$ rises from 0 at $y = 0$ to attain a maximum at $y = \delta$. The function is even about that latter location, which means that for $y > \delta$, $U(y) = U(2\delta - y)$. Similarly, $\langle u'v' \rangle$ is odd about that point. These properties allow us to formulate the problem entirely on the half channel $\{0 \leq y \leq \delta\}$. If the solution is known in the half channel, it can be found by reflection in the other half.

A standard argument shows that the pressure gradient term in the mean momentum balance equation (51) must be constant. Specifically, we refer back to that equation and (52). Setting $\frac{Q}{\rho} = \frac{P}{\rho} - \langle (v')^2 \rangle$, we see that (52) implies that $Q$ depends only on $x$; however by stationarity $\langle (v')^2 \rangle$ does not depend on $x$. The pressure gradient term in (51) can be written $\frac{1}{\rho}\frac{\partial Q}{\partial x}$, which as noted is independent of $y$. It is also independent of $x$, since the other two terms in (51) are.

The friction velocity, Reynolds number $R^*$, small parameter $\epsilon = (R^*)^{-1/2}$, inner scaling (law of the wall) and outer scaling, valid in the channel's center, are all the same as before. Integrating

(51) across the half channel produces a global force balance, resulting in the dimensionless form $\epsilon^2$ for the pressure gradient. The dimensionless form (64) becomes

$$\frac{dT}{d\eta} + 1 + \epsilon^2 \frac{d^2 U}{d\eta^2} = 0, \tag{101}$$

and that of (66) is

$$T + \epsilon^2 \frac{dU^+}{d\eta} = 1 - \eta. \tag{102}$$

The traditional outer approximation is

$$T_{out} = 1 - \eta, \quad 0 < \eta \leq 1. \tag{103}$$

The momentum balance equation with inner normalization is

$$\frac{d^2 U^+}{dy^{+2}} + \frac{dT}{dy^+} + \epsilon^2 = 0. \tag{104}$$

## 7.2 Hierarchy

To exhibit a continuous family of scaling patches covering the channel flow profile, all that is needed is to revise slightly the definition of the adjusted Reynolds stresses (72). The new one is

$$T^\beta(y^+) = T(y^+) + \epsilon^2 y^+ - \beta y^+. \tag{105}$$

It is remarkable that the mathematical problems for the mean velocity and Reynolds stress in these two scenarios—pressure gradient driven and shear driven—can be almost completely transformed one into the other by such a simple device as (105). It transforms the basic momentum balance equation (104) into

$$\frac{d^2 U^+}{dy^{+2}} + \frac{dT^\beta}{dy^+} + \beta = 0, \tag{106}$$

which is of the same form as (74).

Therefore, with the newly adjusted Reynolds stresses, the channel flow context is amenable to the balance exchange processes described in Section 6.1, the construction of a continuum of scalings with associated scaling patches $L_\beta$ in Sections 6.1 to 6.3, and (under some assumptions) the derivation of logarithmic-like profiles in Section 6.4.3. The scaling in $L_\beta$ is still given by (78).

The mean profile calculations are given here only under the simplifying approximation $A =$ constant, although analogs of the more general case can be derived. The expressions (78) and (79) are valid in the present setting as well.

As far as estimating the locations of the patches and the profiles, there are some changes.

The relation (90) becomes

$$\frac{dT}{dy^+} = 4A^{-2}(y^+ - C)^{-2} - \epsilon^2. \tag{107}$$

The constant $C$ in (107) can be related to the location $y^+ = y_m^+$ of the maximum of the original unadjusted function $T$. This is because it is required that $\frac{dT}{dy^+} = 0$ at $y^+ = y_m^+$.

We obtain

$$\frac{dT}{dy^+} = 4A^{-2}\left(y^+ - y_m^+ + \frac{2}{A\epsilon}\right)^{-2} - \epsilon^2, \tag{108}$$

or alternatively by eliminating $\epsilon$,

$$\frac{dT}{dy^+} = 4A^{-2}\left((y^+ - C)^{-2} - (y_m^+ - C)^{-2}\right). \tag{109}$$

When one inserts this into (104) and integrates twice, using the requirement $\frac{dU^+}{dy^+} \to 0$ as $y^+ \to \infty$, the result is

$$U(y^+) = \frac{4}{A^2}\ln\left(y^+ - y_m^+ + \frac{2}{A\epsilon}\right) + C_2 \tag{110}$$

for another integration constant $C_2$.

Strictly speaking, there is another required condition, due to the symmetry of the configuration at the centerline: $\frac{dU^+}{dy^+} = 0$ at $y^+ = \epsilon^{-2}$. It cannot be satisfied exactly within the framework of (110), which indicates that the approximation $A$ = constant cannot be exact near the centerline. The expression (110) is only expected to give a good representation of the real profile in some regions away from both the wall and the centerline.

To reiterate, the most we can say theoretically is that this is suggestive of a logarithmic approximation to some segment of the mean velocity profile. More specifically, this is all under the (doubtful) assumption that $A$ is exactly constant. In the case that it is almost constant, one gets a pair of upper and lower bounds as before, valid now for the mean velocity in channel flow for the range of $y^+$ constructed as before.

Note that in the case $\beta = \epsilon^2$, by (105) $T^\beta = T$. The scaling patch in this particular case is traditionally called the "mesolayer", and it occurs near the peak in Reynolds stress, because for this case $T^\beta = T$. The characteristic length in that patch is $O(\epsilon^{-1})$, which is the geometric mean of those in the inner ($O(1)$) and outer ($O(\epsilon^{-2})$) regions.

## 7.3 Behavior near the wall

The construction of the patches given in sections 6.1 and 7.2 by means of a balance exchange has, as a primary ingredient, the fact that as the peak in adjusted Reynolds stress is approached, a region must appear in which all three terms in the mean momentum balance equation, (61), will have the same order of magnitude. This is simply because the gradient $\frac{dT^+}{dy^+}$ approaches 0. There will, of course, be a smaller region encompassing the peak in which the last term on the left of (61) has smaller order of magnitude than the others, because it vanishes at the peak.

A similar phenomenon happens when $y^+ \to 0$, since that same gradient is zero at the wall ($y^+ = 0$) and positive for small values of $y^+ > 0$. The same conclusion may therefore be deduced: in a small region near the wall, all three terms in (61) will have the same order of magnitude. But the argument in Section 6.1 can only partially be continued beyond this stage to produce a patch with different scaling; in fact it is well known (see also the reason given below) that the characteristic length scale arbitrarily near the wall remains the inner scale. The reason the argument is no longer completely valid will now be explained. In addition, the correctly scaled mean momentum balance very near the wall will be derived.

There have been many analytical, empirical, and computational studies of the properties of the near wall region; we mention only [22], as our results fit particularly well with theirs. Our purpose here is to show that a scaling patch exists there, whose derivation and description fits within the framework of the methodology developed here and in our previous papers.

At the wall, additional constraints are imposed on the functions $U^+$ and $T^+$, besides the basic differential equation (61). First of all, the very definition of inner scaling requires an automatic boundary condition $\frac{dU^+}{dy^+}(0) = 1$. The inner scaling was chosen just so that condition holds. Secondly, the no-slip condition requires the boundary conditions $U^+(0) = T^+(0) = \frac{dT^+}{dy^+}(0) = \frac{d^2T^+}{dy^{+2}}(0) = 0$. The first requirement is simply a result of our choice of normalized variables $y^+$ and $U^+$, and is not a statement of any physical constraint. The other boundary conditions result from a physical effect located at the wall. They have no analog at the mesoscaling patch, and constitute the basic reason that the present construction is different from the mesoscale construction.

If one proceeds in the same vein as before on the basis of (75) and (76), the effect of the first boundary condition $\frac{dU^+}{dy^+}(0) = 1$ is that the length scale in that patch is given by $\alpha = 1$. This means that $\hat{y} = y^+$: the length scale in that patch is the same as that with the original inner normalized scaling. This is of course almost a tautology.

But then the rest of the argument, following (61), in which $\alpha$ and $\gamma$ are determined, can no longer be carried out as it stands, since $\alpha$ has already been determined. However, one can proceed after some reformulation of the problem.

At this point, the first substantial difference in method emerges between the derivation of the patches in Sec. 6.1 and the present argument for what we shall call the wall patch. As mentioned, it is allied with the physical no-slip constraint. In both cases, we look for scaled solutions of (74) or its analog (104) in a neighborhood of a maximum or minimum of $T^+$ or $T_\beta$.

In both the case in Sec. 6.1 and the present wall case, a scaling, namely a choice of $\alpha$, $\gamma$, $\lambda$ in (75), (76), is sought which will render the three terms of (74) or (104) of forms which have the same formal order of magnitude. A unique choice is only possible if one of these three factors is specified by other means. In the previous case, empirical data having to do with the velocity increment across the patch, and also the rate of growth of $U^+$ in the hierarchy, was used to motivate selection of the value $\lambda = 1$, while leaving open the additional possibility of other choices leading to different growth rates. This serves to determine the other two factors. In the wall patch, where this argument is not applicable, the definition of the inner scaling requires $\alpha = 1$, and the equality criterion now can be used to determine $\gamma$ and $\lambda$. Namely, evaluating the three terms of (104) under the transformation (75), (76) with $\alpha = 1$ tells us that $\gamma = \lambda = \epsilon^2$. In neither case does this provide the value of $m$, because the derivatives in (74) or (104) annihilate the linear terms. However, the value of $m$ was obtained by other means: through use of (91) and the connection $T$ has with the slope, in the case of patches embedded in the hierarchy, and by means of the boundary condition giving the slope at the wall, in the present case.

The scaled version of (104) in the wall patch is

$$1 + \frac{d^2\hat{U}}{dy^{+2}} + \frac{d\hat{T}}{dy^+} = 0. \tag{111}$$

The individual terms of this equation are known only at $y^+ = 0$; but it provides a linear relation between the two derivatives. This equation is the analog of (79), and is in the form of a balance of three rescaled forces.

In short, there is a scaling patch near the wall, no doubt including the traditional viscous sublayer, in which the inner length scale is correct, but the deviations of the functions $U^+$ and $T^+$ from their linear parts depend on $R^*$ like $\epsilon^2 \approx (R^*)^{-1}$.

Finally, as $y^+$ enters the patch from above, there is a balance exchange from Layer II (in the terminology used in [7]), where the viscous and turbulent forces balance, to Layer I, where the pressure gradient balances the viscous plus turbulence force.

The location of this exchange, and in fact the size of Layer I, is $\leq O(1)$ in wall units, because that is the length scale for the parameterless (111).

In terms of the traditional buffer and logarithmic layer, we surmise that they lie outside Layer I, which can be identified as the viscous sublayer (although at its outer edge the viscous and turbulent forces are equal in order of magnitude). Outside that layer is approximately where the hierarchy begins (say $y^+ \approx 7$), which is also where the traditionally defined buffer layer begins. The logarithmic mean profile approximation associated with the hierarchy, however, does not become valid until distances from the beginning of the hierarchy are sufficient for $A$ in (87) to be approximately constant.

## 8 Other applications and conclusion

The analysis covered in Secs. 6 and 7 has been applied not only to Couette and Poiseuille flow, but also to

- combined Couette-Poiseuille turbulent flow [23];
- favorable pressure gradient boundary layers (Metzger and Fife, in preparation);
- transport of heat through turbulent channel flow [24].

It should again be emphasized that our goal has been not so much to discover numerical values associated with the profiles, but rather to gain theoretical understanding of why important features occur. We look for answers to *why?* in preference to *what?*

The greatest emphasis has been placed on two approaches to this question:

- the classical Izakson-Millikan observation; and
- the search for scaling patches.

Each of these results is based on an assumption that another approximation is valid. The first relies on the standard inner and outer approximations being simultaneously valid somewhere. The second approach is an argument involving matters of similarity and invariance. There is a family of rescalings depending on a parameter $\beta$, applied at $\beta$-dependent locations, which leave the governing rescaled momentum balance equation and associated numerical values of the derivatives appearing in that equation invariant as $\beta$ is varied. The statement then is that another derivative is, in order of magnitude, invariant as well, and in some regions is in fact approximately numerically invariant. In neither case is there a theoretical way to gauge the error of the assumed approximation, or the extent of the region where it is valid. Some qualitative conclusions on the latter issue are provided in the second case.

The various approaches to answering our basic question all contribute a portion of insight. All of them leave unanswered questions, but they add to one another. No one of them should be considered the final word on the subject.

## Acknowledgments

The results in Secs. 6, 7, and 8 were obtained in collaboration with Joe Klewicki, Tie Wei, Meredith Metzger, and Pat McMurtry. I thank them all for their invaluable exchange of ideas.